\documentclass[12pt]{article}
\usepackage{cite}
\usepackage{amsmath,amsfonts,amssymb}
\usepackage{graphicx,color}
\usepackage{epsfig,epsf}
\usepackage{bm}

\usepackage{relsize}

\usepackage[bookmarks, breaklinks, colorlinks,urlcolor=black, citecolor=red, 
linkcolor=blue]{hyperref}

\usepackage{url}
\usepackage{multirow}
\usepackage{slashed}

\usepackage{ulem}
\def\dsp{\displaystyle}
\def\bea{\begin{eqnarray}}
\def\eea{\end{eqnarray}} 
\def\be{\begin{equation}}
\def\ee{\end{equation}} 
\def\nn {\nonumber}
\def \Re{\text{Re}}

\def\gev{\ensuremath{\mathrm{Ge\kern -0.1em V}}}
\def\mev{\ensuremath{\mathrm{Me\kern -0.1em V}}}

\def\abs#1{\left| #1 \right|}

\definecolor{Darkgreen}{RGB}{30,150,30}

\begin{document}

\begin{flushright}
SI-HEP-2021-19\\[1cm]
%\today
\end{flushright}

%%%%%%%%%%%%%%%%%%%%%%%%%%%%%%%%%%%%%%%%%%%%%%%%%%%%%%%

\begin{center}
	
	{\Large\bf 
	Impact of $B \to K \nu \bar \nu$ measurements on beyond \\[1mm] the Standard Model theories}
	\\[10mm]
	{
	Thomas E. Browder$^a$\,\footnote{teb@phys.hawaii.edu}, Nilendra G. Deshpande$^b$\,\footnote{desh@uoregon.edu}, Rusa Mandal$^c$\,\footnote{Rusa.Mandal@uni-siegen.de (corresponding author)} and Rahul Sinha$^{d,e}$\,\footnote{sinha@imsc.res.in}
}
\\[10pt]
$^a$\,{\small\it  Department of Physics and Astronomy, University of Hawaii, Honolulu, HI 96822, USA} \\[2mm]
$^b$\,{\small\it Institute for Fundamental Science, University of Oregon, OR 97403,USA} \\[2mm]
$^c$\,{\small\it Center for Particle Physics Siegen (CPPS),
	Theoretische Physik 1,   \\ Universit\"at Siegen, 57068 Siegen, Germany}\\[2mm]
$^d$\,{\small\it The Institute of Mathematical Sciences, Taramani, Chennai 600113, India}\\[1mm]
$^e$\,{\small\it Homi Bhabha National Institute Training School Complex, Anushakti Nagar, Mumbai 400085, India}

\end{center}
%%%%%%%%%%%%%%%%%%%%%%%%%%%%%%%%%%%%%%%%%%%%%%%%%%%%%%%

\vspace{5mm}

\begin{abstract}

Semileptonic flavor changing neutral current transitions with a pair of neutrinos in the final state are very accurately determined in the standard model (SM) and thus provide an accurate and sensitive probe for physics beyond the SM. Until recently, the poor tagging efficiency for the $B\to K^{(*)}\nu \bar{\nu}$ modes made them less advantageous as a probe of new physics (NP) compared to the charged lepton counterparts. The most recent Belle II result on $B\to K \nu \bar{\nu}$ uses an innovative inclusive tagging technique resulting in a higher tagging efficiency; this together with previous BaBar and Belle results indicates a possible enhancement in the branching fraction of $B^+\to K^+ \nu \bar{\nu}$. A reanalysis of the full Belle dataset together with upcoming Belle II dataset is expected to result in a much more precise measurement of this mode. If the branching ratio is indeed found to be enhanced with improved measurements, this would provide an unambiguous signal of NP without uncertainties due to long-distance non-factorizable effects or power corrections (in contrast to $B\to K^{(*)} \ell \ell$). We have explored the possibilities of such an enhancement as a signal of NP within several scenarios, which can also explain some of the other tensions observed in neutral as well as charged current $B$-decays.  In an effective field theory approach, with the most general dimension-six Hamiltonian including light right-handed neutrinos, we explore the viability of all scalar and vector leptoquarks as well as the parameter space possible with a generic vector gauge boson $Z^\prime$ model assuming minimal new particle content. While being consistent with all data, correlations between the observed intriguing discrepancies in $B$-decays are also obtained, which will discriminate between the various NP scenarios.

\end{abstract}

%%%%%%%%%%%%%%%%%%%%%%%%%%%%%%%%%%%%%%%%%%%%%%%%%%%%%%%%%
\section{Introduction}
\label{sec:Intro}

Flavor changing neutral current (FCNC) decays are expected to play a significant role
in the search for physics beyond the Standard Model (SM) since they are one-loop
suppressed in the SM. Among such decays, transitions involving $b\to s$ decays have been the subject of attention due to the persistent observation of anomalies indicating the possible existence of new physics (NP). The decays $B_{s}\to \mu^+\mu^-$, $B\to K^{(*)}\ell^+\ell^-$ and $B \to K^{(*)}\nu\bar{\nu}$ are examples of such decays. However, these decays differ in their ability to probe the structure of the SM effective Lagrangian as well as the different contributions from NP. The decay $B\to K^{(*)} \ell^+ \ell^-$ includes contributions from all three of the so-called electroweak penguin operators $\mathcal{O}_7$, $\mathcal{O}_9$ and $\mathcal{O}_{10}$ possible within the SM Hamiltonian, whereas the decay $B_{s}\to \mu^+\mu^-$ has contributions only from a single electroweak penguin operator
$\mathcal{O}_{10}$. Of particular interest to our study is the decay $B \to K^{(*)}\nu\bar{\nu}$, which receives contributions from only two of the
electroweak penguin operators $\mathcal{O}_9$ and $\mathcal{O}_{10}$, with the photon pole contribution absent due to the $\nu\bar{\nu}$ leptonic final state. This results in a much cleaner mode to study with almost no theoretical uncertainty. Noticeably absent are the electromagnetic corrections to hadronic operators that result in difficult to calculate non-local contributions~\cite{Khodjamirian:2010vf} as well as the resonance contributions, both of which plague the interpretation of the results on $B\to K^{(*)} \ell^+ \ell^-$ decays.  The estimation of the $B \to K^{(*)}\nu\bar{\nu}$ decay rates depends only on form-factors, which are considered to be reliably predicted within the SM. Hence, any deviation in the observed decay rate would provide an unambiguous signal of NP.

The observation of the decay mode $B \to K^{(*)}\nu\bar{\nu}$ has until recently
suffered from a serious drawback since the neutrinos are undetected making it
difficult to reconstruct the decay. This required an explicit reconstruction of
the $B$-meson used to tag the signal decay in either a hadronic decay or semileptonic
decay. This tagging suppresses the background but results in a very
low signal reconstruction efficiency,  below 1\%. As a result, no signal has
been observed even for the  $B \to K\nu\bar{\nu}$ mode. A recently introduced
innovative and independent inclusive tagging approach promises to bring a vast
improvement to the experimental study of these modes. Belle II has studied $B^+
\to K^+\nu\bar{\nu}$ using an approach that relies on the inclusive reconstruction of the opposite $B$-meson using charged tracks and photons, as well as vertex information, which results in a larger signal efficiency of about 4\%, at the cost of higher background levels~\cite{Abudinen:2021emt}. Given the
anticipated improvement possible with a reanalysis of Belle data using this
technique and further data from Belle II, an observation of $B^+\to
K^+\nu\bar{\nu}$ seems imminent. %If the $B\to K \nu \bar{\nu}$ branching ratio were found to differ from the SM theoretical prediction, it would provide an unambiguous signal of NP that is free from poorly understood long distance hadronic effects. This mode hence provides clean constraints on possible NP contributions. 

Our goal in this paper is to study the impact of improvements in the observation of $B \to K^{(*)}\nu\bar{\nu}$ on a few popular ``simplified" NP models, such as the leptoquark and generic heavy $Z^\prime$ models. In this context, ``simplified" refers to a single new heavy (above the electroweak scale) mediator particle that can be integrated out to contribute to one or more of the effective operators entering into the $b \to s \nu \bar{\nu}$ transition. The choice of these ``simplified" NP models is motivated from the fact that they are the prime candidates capable of explaining the other intriguing hints of NP observed in $B$-decay. 

We can imagine a number of future scenarios. For example, if an enhancement of $B \to K^{(*)}\nu\bar{\nu}$ is found, which is consistent with the apparent NP signatures in $B \to K^{(*)}\ell \ell$ and/or $B\to D^{(*)} \tau \nu$, it would provide a compelling case for NP. Another possibility is that with improved experimental precision, the branching fraction for $B \to K^{(*)}\nu\bar{\nu}$ approaches the SM expectation but the anomalies in the other channels persist. This would then lead to strong constraints in the NP model parameter space.

In addition to possible anticipated improvements in $B \to K^{(*)}\nu\bar{\nu}$ branching ratios we consider several other observables that also impact the NP parameter space.  These include, in the neutral current (NC) channels,  the lepton flavor universality ratios $R_K$~\cite{Aaij:2019wad,Aaij:2021vac} and $R_{K^*}$~\cite{Aaij:2017vbb}; a full set of angular observables for the modes $B^+ \to K^+ \mu^+ \mu^-$~\cite{Aaij:2012vr}, $B \to K^{*} \mu^+ \mu^-$~\cite{Aaij:2020nrf,Aaij:2020ruw} and $B_s^0 \to \phi \mu^+ \mu^-$~\cite{Aaij:2015esa}; the branching ratio $B_s \to \mu^+\mu^-$~\cite{Bsmumu,CMS:2020rox,LHCb:2021vsc}; and $B_s$ mixing data~\cite{Aaij:2019vot}. From the point of view of the charged current (CC) transition, the important observables are the lepton flavor universality ratios $R(D^{(*)})$~\cite{Amhis:2019ckw}. Among other observables, electroweak precision measurements~\cite{Zyla:2020zbs} are also critical in some cases. There have been previous studies~\cite{Buras:2014fpa,Kamenik:2011vy,Bobeth:2001jm,Altmannshofer:2009ma} analyzing the effect of NP models in $B \to K^{(*)}\nu \bar{\nu}$ modes, some focusing on the connection with $b\to s \mu \mu$ anomalies in an effective theory approach where the flavor structure is dictated by the assumption of minimal flavor violation~\cite{Descotes-Genon:2020buf}. However, in this work we explore the NP parameter space connecting both $b\to s \mu \mu$ and $b \to c \tau \bar{\nu}$ tensions and also include light right-handed neutrinos (RHNs). Thus, the computations of these NP contributions are performed in the Standard Model Effective Theory (SMEFT), as well as in the neutrino-Weak Effective Theory ($\nu$-WET) basis. Instead of requiring these NP models to just fulfill the bounds on $b\to s \nu \bar{\nu}$ channels, we analyze whether the anticipated signal can bear any footprint on the discrepancies observed in NC and CC $B$-decays.

The paper is organized as follows:  Starting with the most general dimension-6 effective Hamiltonian (including RHNs), in Sec.~\ref{sec:theory}, we derive the relevant observables and study the individual effects of Wilson coefficients.  Section~\ref{sec:obs} deals with the NC and CC observables and the matching to SMEFT basis for the four fermion operators. We discuss several scenarios with scalar as well as vector leptoquarks in Sec.~\ref{sec:LQs} and a generic $Z^\prime$ setup in Sec.~\ref{sec:Z'}, and then discuss the outcomes connecting $B$-anomalies and $B \to K^{(*)}\nu\bar{\nu}$ in Sec.~\ref{sec:dis}. Our concluding remarks are given in Sec.~\ref{sec:summary}.

%%%%%%%%%%%%%%%%%%%%%%%%%%%%
\section{Theoretical framework}
\label{sec:theory}
%%%%%%%%%%%%%%%%%%%%%%%%%%%%

Including light RHN fields, the
most general dimension-6 effective Hamiltonian relevant for $b\to s \nu^\alpha \bar\nu^\beta$  transitions can be written, at the bottom quark-mass scale, as
\begin{equation}
\label{eq:Heff}
{\cal H}_\text{eff}\, =\, -\frac{ 4 G_F  }{\sqrt{2}} \frac{\alpha_\text{EM}}{4\pi} V_{tb}V_{ts}^*  \left( C_{LL}^{\rm SM} \delta_{\alpha \beta} [{\cal O}_{LL}^V]^{\alpha \beta} + \sum_{\substack{X=S,V,T \\ A,B=L,R}} [C_{AB}^X]^{\alpha \beta} [{\cal O}_{AB}^X]^{\alpha\beta} \right)\,,% (\bar{s}\gamma^\mu P_A b)(\bar{\nu}^{\alpha} \gamma_\mu P_B \nu^{\beta})\,.
\end{equation}
with the ten four-fermion operators:
\begin{eqnarray}\label{eq:K-operators}
[{\cal O}^V_{AB}]^{\alpha\beta} &\equiv & \left(\bar{s}\, \gamma^\mu P_A b\right)\left(\bar{\nu}^{\alpha} \gamma_\mu P_B \nu^\beta\right)\, ,
\nn\\
{[{\cal O}^{S}_{AB}]}^{\alpha\beta} &\equiv & \left(\bar{s}\, P_A b\right)\left(\bar{\nu}^{\alpha} P_B \nu^\beta\right)\, ,
\nn\\
{[{\cal O}^T_{AB}]}^{\alpha\beta} & \equiv & \delta_{AB}\;\left(\bar{s}\,\sigma^{\mu \nu} P_A b\right)\left(\bar{\nu}^{\alpha} \sigma_{\mu \nu} P_B \nu^\beta\right)\, ,
\end{eqnarray}
The SM FCNC contribution to $[{\cal O}^V_{AB}]^{\alpha\alpha} $ has been explicitly added to Eq.~\eqref{eq:Heff}, where the Wilson coefficient~\cite{Buras:2014fpa},
\bea
C_{LL}^{\rm SM}= - 2 X_t/s_w^2\, \quad {\rm with} \quad X_t= 1.469 \pm 0.017\,,
\eea 
includes NLO QCD corrections and two-loop electroweak contributions.
All other Wilson coefficients $[C_{AB}^X]^{\alpha \beta}=0$ (except for a negligible contribution to $[C_{RL}^V]^{\alpha \alpha}$) in the SM, and thus any nonzero contribution to these Wilson coefficients, is then a manifestation of NP beyond the SM.

The differential branching fraction with respect to the dineutrino invariant mass squared ($q^2$) for $B\to K^{(*)}\nu\bar\nu$ decays with the effective Hamiltonian in Eq.~\eqref{eq:Heff} can be written as
\begin{align}
\label{eq:Kdist}
\frac{d\Gamma}{dq^2}(B \to K \nu \bar{\nu})\; &=\;  \frac{G_F^2 |V_{tb} V_{ts}^*|^2 \alpha_\text{EM}^2}{192\times  16  \pi^5 m_B^3}\;  q^2 \, \lambda_K^{1/2} (q^2)\times \nn \\
&\times\!\!   \sum_{\alpha=1}^3  \sum_{\beta=1}^3 \Big[  \Bigl(	 |C_{LL}^{\rm SM} \delta_{\alpha \beta}+[C^V_{LL}]^{\alpha\beta}+[C^V_{RL}]^{\alpha\beta}|^2+ |[C^V_{LR}]^{\alpha\beta}+[C^V_{RR}]^{\alpha\beta}|^2  	\Bigr)	(H^{s}_{V})^2  
 \nn \\
& \qquad \qquad +\frac{3}{2}\,  \Bigl( |[C^S_{RL}]^{\alpha\beta}+[C^S_{LL}]^{\alpha\beta}|^2 +|[C^S_{RR}]^{\alpha\beta}+[C^S_{LR}]^{\alpha\beta}|^2 \Bigr) (H^s_S)^2  \nn \\
& \qquad \qquad+8\, \Bigl( |[C^T_{LL}]^{\alpha\beta}|^2 + |[C^T_{RR}]^{\alpha\beta}|^2\Bigr) (H^s_T)^2 \Big] \,,  
\end{align}
\begin{align}
\label{eq:Kstdist}
\frac{d\Gamma}{dq^2}(B\to K^* \bar\nu\nu) &= \frac{G_F^2 |V_{tb} V_{ts}^*|^2 \alpha_\text{EM}^2}{192\times  16  \pi^5 m_B^3}\;  q^2 \, \lambda_{K^*}^{1/2} (q^2)\times
\nn \\ \nn &
\times\, \sum_{\alpha=1}^3 \sum_{\beta=1}^3    \abs{C_{LL}^{\rm SM}\delta_{\alpha \beta} +[C^V_{LL}]^{\alpha\beta}}^2  \left(H_{V,+}^2+H_{V,-}^2 \right) 
\\ \nn &  \hspace*{2cm}
 +\,\abs{C_{LL}^{\rm SM} \delta_{\alpha \beta}
 	+[C^V_{LL}]^{\alpha\beta}-[C^V_{RL}]^{\alpha\beta}}^2 H_{V,0}^2  
\\ \nn & \hspace*{2cm}
-\, 4\, \Re \left [ \left( C_{LL}^{\rm SM}\delta_{\alpha \beta} + [C^{V}_{LL}]^{\alpha\beta} \right) [C^{V*}_{RL}]^{\alpha\beta} \right]   H_{V, +} H_{V, -}
\\ \nn &  \hspace*{2cm}
+\,   \left( \abs{[C^V_{RL}]^{\alpha\beta}}^2 + \abs{[C^V_{LR}]^{\alpha\beta}}^2 + \abs{[C^V_{RR}]^{\alpha\beta}}^2 \right)  \left(H_{V,+}^2+H_{V,-}^2 \right) 
\nn \\ & \hspace*{2cm}
+ \abs{[C^V_{LR}]^{\alpha\beta}-[C^V_{RR}]^{\alpha\beta}}^2  H_{V,0}^2  - 4 \Re \left[ [C^V_{LR}]^{\alpha\beta} \, [C^{V*}_{RR}]^{\alpha\beta} \right]   H_{V, +} H_{V, -} \nn \\
&  \hspace*{2cm} +\frac{3}{2}\,\left( \abs{[C^{S}_{RL}]^{\alpha\beta} - [C^{S}_{LL}]^{\alpha\beta}}^2 + \abs{[C^{S}_{RR}]^{\alpha\beta} - [C^{S}_{LR}]^{\alpha\beta}}^2 \right) H_S^2 
\nn \\  &  \hspace*{2cm}
+\, 8\, \left( \abs{[C^T_{LL}]^{\alpha\beta}}^2 + \abs{[C^T_{RR}]^{\alpha\beta}}^2 \right)  \left( H_{T,+}^2  + H_{T, -}^2 + H_{T, 0}^2 \right)  \, .
\end{align}
Here we have introduced the shorthand notation for the K\"{a}llen function: 
\be 
\lambda_{K^{(*)}}(q^2) \equiv \lambda(m_B^2,m_{K^{(*)}}^2, q^2)=m_B^4+ m_{K^{(*)}}^4+ q^4 - 2 m_B^2m_{K^{(*)}}^2 - 2 m_{K^{(*)}}^2 q^2- 2 m_B^2 \, q^2\,.
\ee
The expressions for the helicity amplitudes in terms of the form factors are given in Appendix~\ref{sec:FF}, and for notational simplicity, we have dropped the explicit $q^2$ dependence in helicity amplitudes here. It is important to mention that the NP operators introduce new helicity amiplitudes ($H_{S,T}^s$ for the $K$ mode; $H_S$ and $H_{T,\lambda}$ with $\lambda=+,\,-,\,0$ for the $K^*$ channel), which depend on extra form factors. This gives rise to an extra source of uncertainty in the observables; however, the predictions for these form factors are accurately known from lattice QCD~\cite{Bailey:2015dka} and light-cone sum rule computations~\cite{Straub:2015ica}. In the context of beyond the SM theories with light NP states such as RHNs or dark matter particles, the $q^2$ variation of these differential distributions will be important in order to discriminate between different scenarios. In the rest of our analysis, we assume the $\nu_R$ fields to be light,	$m_{\nu_R} \lesssim \mathcal{O}(100)$\,MeV, so that they do not modify the differential distributions of $B\to K^{(*)}\nu\bar{\nu}$.

It is easy to see that the expressions in Eqs.~\eqref{eq:Kdist} and \eqref{eq:Kstdist} encode all possible flavor diagonal as well as off-diagonal neutrino contributions. In the absence of scalar and tensor operators the branching ratio for $B\to K\nu\bar\nu$ simplifies to
\begin{align}
\label{eq:KBR}
\mathcal{B}(B\to K \bar\nu\nu) = \mathcal{B}(B\to K\bar\nu\nu)|_\text{SM} \times & \Bigg[ \frac{1}{3}\,\sum_{\alpha=1}^3  \bigg| 1 +\frac{C_{LL}^{\alpha\alpha}+ C_{RL}^{\alpha\alpha}}{C_{LL}^\text{SM}}\bigg|^2 + \sum_{\alpha\neq\beta} \bigg| \frac{C_{LL}^{\alpha\beta}+ C_{RL}^{\alpha\beta}}{C_{LL}^\text{SM}} \bigg|^2  \nn \\
&+\sum_{\alpha=1}^3 \sum_{\beta=1}^3 \bigg|\frac{C_{LR}^{\alpha\beta} + C_{RR}^{\alpha\beta}}{C_{LL}^\text{SM}}\bigg|^2\Bigg]\,.
\end{align}

In the case of $B\to K^* \nu \bar{\nu}$, another observable can be constructed: namely, the longitudinal polarization fraction of $K^*$, which in the presence of all possible dimension-6 operators quoted in Eq.~\eqref{eq:Heff}, can be written as
\begin{align}
F_L^{K^*\nu\bar{\nu}}(q^2)= &\frac{G_F^2 |V_{tb} V_{ts}^*|^2 \alpha_\text{EM}^2}{192\times  16  \pi^5 m_B^3}\;  q^2 \, \lambda_{K^*}^{1/2} (q^2) \frac{1}{d\Gamma(B\to K^* \bar\nu\nu)/dq^2} \times \nn \\
&\sum_{\alpha=1}^3 \sum_{\beta=1}^3 \big[ \abs{C_{LL}^{\rm SM} \delta_{\alpha \beta} +[C^V_{LL}]^{\alpha\beta}-[C^V_{RL}]^{\alpha\beta}}^2 H_{V,0}^2 -8 |C_{LL}^T|^2 H_{T,0}^2  \nn \\
&\hspace*{1.5cm}+ \abs{[C^V_{LR}]^{\alpha\beta}-[C^V_{RR}]^{\alpha\beta}}^2 H_{V,0}^2 -8 |C_{RR}^T|^2 H_{T,0}^2\big]\,.
\end{align}

\begin{figure}[t!]
	\centering
	\includegraphics[scale=0.55]{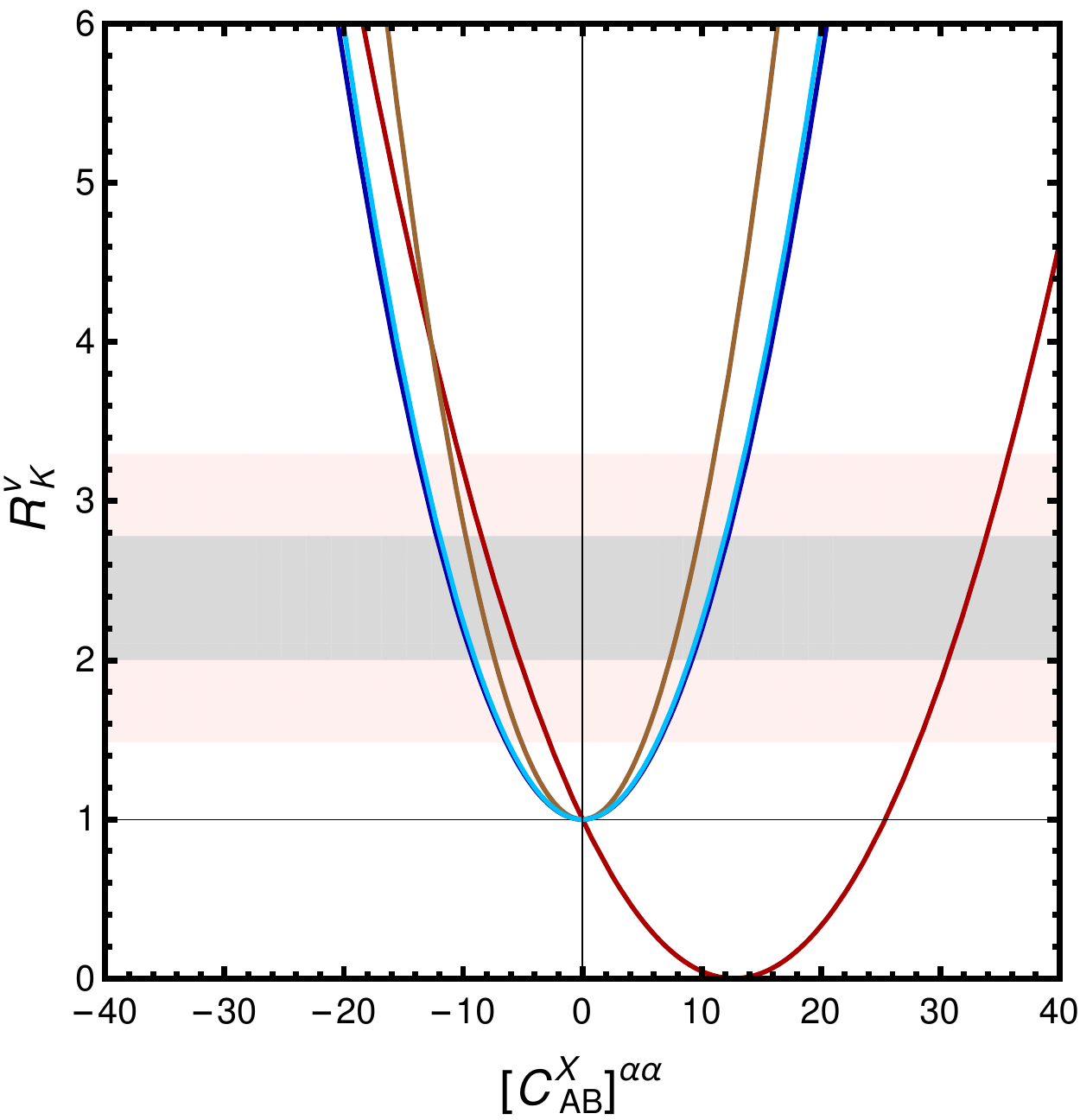}\hskip 30pt
	\includegraphics[scale=0.55]{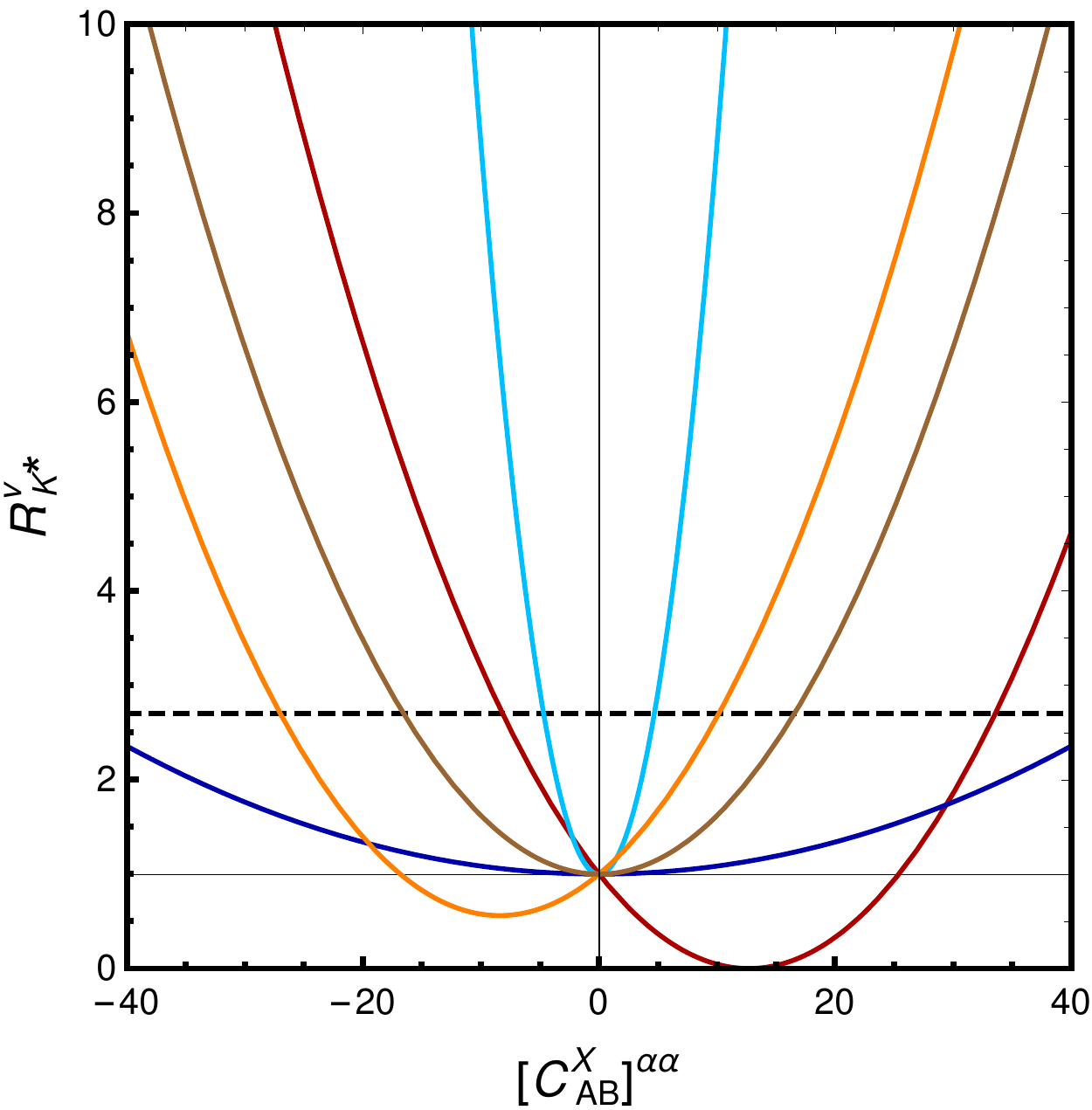}\vskip 10pt
	\includegraphics[scale=0.87]{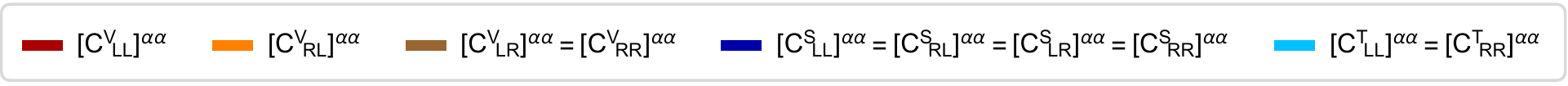}
	\caption{Variations of flavor diagonal individual NP Wilson coefficients (assuming the same contribution for all three generations) are shown for the observables $R_K^\nu$ and $R_{K^*}^\nu$ in the left and right panels, respectively. The red and gray bands in the left panel show the $\pm1\,\sigma$ signal strength quoted in Eq.~\eqref{eq:RKnu} and the future expectation (uncertainties reduced by a factor of three), respectively, whereas the  black dashed line in the right panel is the upper bound given in Eq.~\eqref{eq:RKstnu}.}
	\label{fig:var}
\end{figure}

The SM expectations for the observables in $B\to K^{(*)}\nu\bar\nu$ decays are calculated in Refs.~\cite{Blake:2016olu,Buras:2014fpa} and read as follows:
\begin{align}
\label{eq:B2KSM}
\mathcal{B}(B\to K^{+}\bar\nu\nu)\vert_\text{SM} =&  (4.6\pm0.5)\times 10^{-6} \,,\\
\mathcal{B}(B\to K^{*0}\bar\nu\nu)\vert_\text{SM} =&  (8.4\pm1.5)\times 10^{-6}\,,\\
 F_L^{K^*\nu\bar{\nu}} |_{\rm SM} =& 0.47\pm 0.03\,.
\end{align}
The experimental upper limits for these modes are
\begin{align}
\mathcal{B}(B\to K^{+}\bar\nu\nu) <& \,1.6\times 10^{-5} \,\text{\cite{Grygier:2017tzo}}\,, \\
\mathcal{B}(B\to K^{*0}\bar\nu\nu) <& \, 2.7\times 10^{-5} \, \text{\cite{Grygier:2017tzo}} \,,\\
\mathcal{B}(B\to K^{*+}\bar\nu\nu) <&\, 4.0\times 10^{-5}\, \text{\cite{Lees:2013kla}}\,,
\end{align}
where all three results are obtained at the 90\% C.L., and so far no data exists for $ F_L^{K^*\nu\bar{\nu}}$. Recently using the inclusive tag technique, the Belle II collaboration has reported a signal strength of $\mu=4.2^{+3.4}_{-3.2} $ ~\cite{Abudinen:2021emt} with an initial 63\,fb$^{-1}$ of data sample, which when combined with previous measurements by Belle~\cite{Grygier:2017tzo,Lutz:2013ftz} and Babar~\cite{Lees:2013kla} gives rise to a world average for the branching fraction of $(1.1 \pm 0.4)\times 10^{-5}$~\cite{Dattola:2021cmw}. Now, including the SM prediction (in Eq.~\eqref{eq:B2KSM}), we obtain for the ratio
\bea
\label{eq:RKnu}
R_K^\nu \equiv \frac{\mathcal{B}(B\to K^{+}\bar\nu\nu)}{\mathcal{B}(B\to K^{+}\bar\nu\nu)|_{\rm SM}} = 2.4 \pm 0.9\,.
\eea
The analysis with this inclusive tagging technique on the full Belle data is expected to reduce the uncertainty significantly and allow sensitivity to a branching fraction in the range $[1,2]\times 10^{-5}$, consistent with the current world average central value. An expectation with the upcoming Belle II and full Belle datasets would be that the central value remains at the value of Eq.~\eqref{eq:RKnu}, and the uncertainties are reduced by a factor of 3: i.e., $R_K^\nu=2.4\pm 0.3$. In the case of the vector meson counterpart, no signal has been observed so far, and the limit reads
\bea
\label{eq:RKstnu}
R_{K^*}^\nu \equiv \frac{\mathcal{B}(B\to K^{*}\bar\nu\nu)}{\mathcal{B}(B\to K^{*}\bar\nu\nu)|_{\rm SM}} <2.7\,.
\eea
In Fig.~\ref{fig:var}, we show the variation of individual Wilson coefficients considering only diagonal flavor structures. The $1\sigma$ signal region is easily achievable for $R_K^\nu$ for all Wilson coefficients except for some positive values of $[C_{LL}^V]^{\alpha \alpha}$. It can be seen from Eqs.~\eqref{eq:Kdist} and \eqref{eq:Kstdist} that the expressions are symmetric under the interchange of $[C_{LR}^V]^{\alpha \beta} \leftrightarrow [C_{RR}^V]^{\alpha \beta}$, all four scalar operators and two tensor operators. 
The $B \to K \nu \bar{\nu}$ mode is also symmetric under $[C_{LL}^V]^{\alpha \beta} \leftrightarrow [C_{RL}^V]^{\alpha \beta}$, which does not hold for the $K^*$ mode due to interference with different helicity structures.

Focusing on a SMEFT approach in which all SM fields are dynamical degrees of freedom and only the heavy NP is integrated out, we perform a matching of the Wilson coefficients of the effective Hamiltonian in Eq.~\eqref{eq:Heff} contributing to $b \to s \nu^\alpha \bar \nu^\beta$ in terms of the four fermion operators written in the Warsaw basis ~\cite{Grzadkowski:2010es}  (quoted in Appendix~\ref{sec:SMEFT})  as
\begin{align}
[C_{LL}^V]^{\alpha\beta} = & \,+\frac{2v^2}{\Lambda^2}\frac{\pi}{\alpha_\text{EM}|V_{tb}V^*_{ts}|}\left([\mathcal{C}_{lq}^{(1)}]^{23\alpha\beta}-[\mathcal{C}_{lq}^{(3)}]^{23\alpha\beta}\right) \,,\\
[C_{RL}^V]^{\alpha\beta} = & \,+\frac{2v^2}{\Lambda^2}\frac{\pi}{\alpha_\text{EM}|V_{tb}V^*_{ts}|} [\mathcal{C}_{ld}]^{23\alpha\beta} \,.
\label{eq:matching_BKnunu}
\end{align}

\section{Observables}
\label{sec:obs}
Here we briefly mention the list of observables in which certain tensions have been observed between data and the SM theory expectations: namely, the NC mode $b\to s \ell\ell$ and the CC $b\to c \tau \bar{\nu}$ transition. Our main aim is to correlate these tensions with the $b \to s \nu \bar \nu$ channels: however, there exist several other observables that impose strong constraints on the NP parameter space and will be discussed in the corresponding context below.

\subsection{$b\to s \ell^+ \ell^-$}
\label{app:dtodll}
The effective Hamiltonian describing the $b\to s \ell^\alpha\ell^\beta$ FCNC transition is
\be
\begin{aligned}
	\mathcal{H}^\text{eff}(b\to s \ell^\alpha\ell^\beta)=-\frac{4 G_F}{\sqrt{2}}\frac{\alpha_\text{EM}}{4\pi}V_{tb}V^*_{ts}&\left[C_9^{(\prime)\,\alpha\beta}\mathcal{O}_9^{(\prime)\,\alpha\beta} +C_{10}^{(\prime)\,\alpha\beta}\mathcal{O}_{10}^{(\prime)\,\alpha\beta} - \frac{2 m_b }{q^2}\,C_{7}^{(\prime)\,\alpha\alpha}\mathcal{O}_{7}^{\alpha\alpha} \right. \\ 
	&+ \left.    C_S^{(\prime)\,\alpha\beta} \mathcal{O}_S^{(\prime)\,\alpha\beta} +   C_{P}^{(\prime)\,\alpha\beta} \mathcal{O}_P^{(\prime)\,\alpha\beta}  \right ]\,,
\end{aligned}
\label{eq:lagrangian_bsll}
\ee
where the operators are
\begin{align}
\mathcal{O}_9^{(\prime)\,\alpha\beta} &= (\bar{s} \gamma^\mu  P_{L(R)} b)(\bar{\ell}^\alpha \gamma_\mu \ell^\beta)  \,,&\mathcal{O}_S^{(\prime)\,\alpha\beta} &=  \,(\bar{s}  P_{R(L)} b)(\bar{\ell}^\alpha \ell^\beta)  \,,\\ ~\mathcal{O}_{10}^{(\prime)\,\alpha\beta} & = (\bar s \gamma^\mu P_{L(R)} b)(\bar{\ell}^\alpha \gamma_\mu\gamma_5 \ell^\beta)\,,&
\mathcal{O}_P^{(\prime)\,\alpha\beta} &=  \,(\bar{s}  P_{R(L)} b)(\bar{\ell}^\alpha \gamma_5\ell^\beta)  \,,\\
\mathcal{O}_{7}^{(\prime)\,\alpha\alpha} &=(\bar s i \sigma^{\mu\nu} q_\nu P_{R(L)} b)(\bar{\ell}^\alpha \gamma_\mu \ell^\alpha)\,.
\label{eq:kstllOp}
\end{align}
The matching to SMEFT Wilson coefficients quoted in Appendix~\ref{sec:SMEFT} gives~\cite{Alonso:2014csa}
\be
\begin{aligned}
	C_{9}^{\alpha\beta} &= C_9^{ \text{SM}}\delta_{\alpha\beta}  -\frac{v^2}{\Lambda^2}\frac{\pi}{\alpha_\text{EM} V_{tb}V^*_{ts}} \left([\mathcal{C}_{lq}^{(3)}]^{23\alpha\beta}+[\mathcal{C}_{lq}^{(1)}]^{23\alpha\beta}+ [\mathcal{C}_{eq}]^{23\alpha\beta}\right)\,, \\
	C_{10}^{\alpha\beta} &= C_{10}^{ \text{SM}}\delta_{\alpha\beta}  +\frac{v^2}{\Lambda^2}\frac{\pi}{\alpha_\text{EM} V_{tb}V^*_{ts}} \left([\mathcal{C}_{lq}^{(3)}]^{23\alpha\beta}+[\mathcal{C}_{lq}^{(1)}]^{23\alpha\beta}- [\mathcal{C}_{eq}]^{23\alpha\beta}\right)\,, \\
	C_{9}^{\prime\, \alpha\beta} &=  -\frac{v^2}{\Lambda^2}\frac{\pi}{\alpha_\text{EM} V_{tb}V^*_{ts}} \left([\mathcal{C}_{ld}]^{23\alpha\beta}+[\mathcal{C}_{ed}]^{23\alpha\beta}\right)\,, \\
	C_{10}^{\prime\, \alpha\beta} &=  +\frac{v^2}{\Lambda^2}\frac{\pi}{\alpha_\text{EM} V_{tb}V^*_{ts}} \left([\mathcal{C}_{ld}]^{23\alpha\beta}-[\mathcal{C}_{ed}]^{23\alpha\beta} \right)\,, \\
	C_S^{\alpha \beta} &= - C_P^{\alpha \beta} = \frac{v^2}{\Lambda^2}\frac{\pi}{\alpha_\text{EM} V_{tb}V^*_{ts}} [\mathcal{C}_{ledq}^*]^{23\alpha\beta} \,,\\
	C_S^{\prime\,\alpha \beta} &= C_P^{\prime\,\alpha \beta} = \frac{v^2}{\Lambda^2}\frac{\pi}{\alpha_\text{EM} V_{tb}V^*_{ts}} [\mathcal{C}_{ledq}]^{32\alpha\beta}\,,
\end{aligned}
\label{eq:WCs_bsll}
\ee
where $C_9^{\rm SM}(m_b)= 4.211,~C_{10}^{\rm SM}(m_b)= -4.103,~C_7(m_b)=C_7^{\rm SM}=- 0.304$. A common notation separates out the NP contributions in the Wilson coefficients $C_{9,10}$ for the $b\to s \mu \mu$ mode, which are defined as $C_{9,10}^{\rm NP}\equiv C_{9,10}^{22}- C_{9,10}^{\rm SM}$.

The most interesting measurements in these transitions are the lepton flavor universality violating ratios $R_K$~\cite{Aaij:2019wad,Aaij:2021vac} and $R_{K^*}$~\cite{Aaij:2017vbb}, which have tensions when compared with the corresponding SM predictions. There also exist branching fraction measurements for the purely leptonic decay $B_s \to \mu^+\mu^-$~\cite{CMS:2020rox,Bsmumu,LHCb:2021vsc} and the full set of angular observables for the modes $B^+ \to K^+ \mu^+ \mu^-$~\cite{Aaij:2012vr}, $B^0 \to K^{*0} \mu^+ \mu^-$~\cite{Aaij:2020nrf}, and $B_s^0 \to \phi \mu^+ \mu^-$~\cite{Aaij:2015esa} with the very recently measured $B^+ \to K^{*+} \mu^+ \mu^-$~\cite{Aaij:2020ruw}.
In order to explain the observed tensions, a global fit to all available data in the $b\to s \mu \mu$ channel provides preferred intervals for the Wilson coefficients~\cite{Altmannshofer:2021qrr,Geng:2021nhg,Hurth:2021nsi,Alguero:2021anc,Carvunis:2021jga}. Since global fits to constrain NP involving $B\to K^*\mu\mu$ angular observables are based on ignoring possible hadronic contributions, it is a basic assumption that these effects are small, and any tensions arise only due to NP. This holds true even for the NP analysis done in this paper.

\subsection{$b \to c \ell \bar \nu$}

The charged-current transition $b\to c \ell_\alpha\bar\nu_\beta$ is described by the following Hamiltonian
\begin{equation}
\label{eq:Heffb2c}
{\cal H}_\text{eff}\, =\, \frac{ 4 G_F V_{cb}}{\sqrt{2}}\left( {\cal Q}_{LL}^{V\alpha\beta} \delta_{\alpha\beta} + \sum_{\substack{X=S,V,T \\ 
		A,B=L,R}} \mathcal{P}_{AB}^{X\alpha\beta}\; {\cal Q}_{AB}^{X\alpha\beta} \right),
\end{equation}
with the ten four-fermion operators:
\begin{eqnarray}\label{eq:def-operators}
{\cal Q}^{V\alpha\beta}_{AB} &\equiv & \left(\bar{c}\, \gamma^\mu P_A b\right)\left(\bar{\ell}^\alpha \gamma_\mu P_B \nu^\beta \right)\, ,
\nn\\
{\cal Q}^{S\alpha\beta}_{AB} &\equiv & \left(\bar{c}\, P_A b\right)\left(\bar{\ell}^\alpha P_B \nu^\beta \right)\, ,
\nn\\
{\cal Q}^{T\alpha\beta}_{AB} & \equiv & \delta_{AB}\;\left(\bar{c}\,\sigma^{\mu \nu} P_A b\right)\left(\bar{\ell}^\alpha \sigma_{\mu \nu} P_B \nu^\beta\right)\, .
\end{eqnarray}
The NP Wilson coefficients mapped to the SMEFT basis of Appendix~\ref{sec:SMEFT} are found to be
\begin{align}
\label{eq:btoc_coeffs}
\mathcal{P}_{LL}^{V\alpha\beta}=&+ \frac{v^2}{\Lambda^2}\, \sum_{m=1}^3 \frac{V_{2m}}{V_{cb}} [\mathcal{C}^{(3)}_{lq}]^{m3\alpha\beta}  \,,  \\
\mathcal{P}_{LL}^{S\alpha\beta} =&-\frac{v^2}{2 \Lambda^2 V_{cb}}\,[\mathcal{C}^{(1)*}_{lequ}]^{23\alpha\beta}\,,   \\
\mathcal{P}_{LL}^{T\alpha\beta}=&-\frac{v^2}{2 \Lambda^2 V_{cb}} [\mathcal{C}^{(3)*}_{lequ}]^{23\alpha\beta}\,, \\
\mathcal{P}_{RL}^{S\alpha\beta} =&+\frac{v^2}{2 \Lambda^2}\, \sum_{m=1}^3 \frac{V_{2m}}{V_{cb}}[\mathcal{C}^{*}_{ledq}]^{m3\alpha\beta}\,. 
\end{align}
 In this case, among the observables the most relevant ones are the lepton flavor universality violating ratios $R(D)$ and $R(D^*)$~\cite{Amhis:2019ckw} and the longitudinal helicity fraction for $D^*$~\cite{Abdesselam:2019wbt}. The $\tau$ polarization measurement~\cite{Hirose:2016wfn} in $B \to D^* \tau \bar \nu$ and the differential branching fractions for both $D$ and $D^*$ data have large uncertainties~\cite{Huschle:2015rga,Lees:2013uzd}. The expressions for the relevant observables in terms of the Wilson coefficients are quoted in Appendix~\ref{sec:CCexps}.

\section{New Physics Models}
\label{sec:models}

In this section, we discuss two popular NP scenarios: namely, leptoqarks and a generic $Z^\prime$ in a ``simplified" NP model setup. Only one mediator is assumed to be present at a time and it contributes to the $b\to s \nu \bar{\nu}$ channel at tree level.

\begin{table}[ht!]
	\centering
	\renewcommand\arraystretch{1.2}
	\resizebox{1.\textwidth}{!}{\hspace*{-1.cm}
	\begin{tabular}{c| c|  c|  c|  c }
		\hline\hline
		Spin & Leptoquarks  & Interaction  &  SMEFT  operators & $\nu$-WET operators \\
		& ($\mathcal{G}_{\rm SM}$)&terms $(\mathcal{L}_{\rm NP})$ & &  \\
		\hline\hline \noalign{\vskip 2pt}
		0 & $S_3(\bar{\bf 3},{\bf 3},1/3)$ & $+\,\overline{Q^c}\, Y_{\tiny S_3} \,i\tau_2  \, {\boldsymbol \tau\bf \cdot S_3} \, L$ &  $\dsp\frac{[\mathcal{C}_{lq}^{(1)}]^{ij\alpha \beta}}{\Lambda^2}= \dsp\frac{3[\mathcal{C}_{lq}^{(3)}]^{ij\alpha \beta}}{\Lambda^2}= - \frac{3Y_{\tiny S_3}^{j\beta} Y_{\tiny S_3}^{*i\alpha} }{4M_{\tiny S_3}^2}$ &  $[N_{LL}^{V,d}]^{ij\alpha \beta}=  \dsp\frac{Y_{\tiny S_3}^{j\beta} Y_{\tiny S_3}^{*i\alpha} }{2M_{\tiny S_3}^2}$  \\[2ex]
		\hline
		%%%%%%%%%%%%%%%%%%%%%%%%%%%%%%%%%%
			\multirow{4}{*}{0}  & 	\multirow{4}{*}{$\tilde{R}_2({\bf 3},{\bf 2},1/6)$} & 	\multirow{4}{*}{$-\,\overline{d}_R  \,Y_{\tiny \tilde{R_2}}\,  \tilde{R}_2^T\, i \tau_2 L$ }&  %
			\multirow{4}{*}{$\dsp\frac{[\mathcal{C}_{ld}]^{ij\alpha \beta}}{\Lambda^2}= - \frac{Y_{\tiny \tilde{R}_2}^{i\beta} Y_{\tiny \tilde{R}_2}^{*j\alpha} }{2M_{\tiny \tilde{R}_2}^2}$} & 	\multirow{4}{*}{ $[N_{RL}^{V,d}]^{ij\alpha\beta}= -\dsp \frac{Y_{\tiny \tilde{R}_2}^{i\beta} Y_{\tiny \tilde{R}_2}^{*j\alpha} }{2M_{\tiny \tilde{R}_2}^2}$} \\[7ex] 
			&& +\,$\overline{Q}  \,Z_{\tiny \tilde{R}_2}\,  \tilde{R}_2\, \nu_R$ & &$[N_{LR}^{V,d}]^{ij\alpha\beta}= -\dsp \frac{Z_{\tiny \tilde{R}_2}^{i\beta} Z_{\tiny \tilde{R}_2}^{*j\alpha} }{2M_{\tiny \tilde{R}_2}^2}$ \\[2ex]
			&& &  & $[N_{LL}^{S,d}]^{ij\alpha\beta}= 4[N_{LL}^{T,d}]^{ij\alpha\beta}= \dsp -\frac{Y_{\tiny \tilde{R}_2}^{i\beta} Z_{\tiny \tilde{R}_2}^{*j\alpha} }{2M_{\tiny \tilde{R}_2}^2}$ \\[2ex] 
			&&  & & $[N_{RR}^{S,d}]^{ij\alpha\beta}= 4[N_{RR}^{T,d}]^{ij\alpha\beta}= \dsp -\frac{Z_{\tiny \tilde{R}_2}^{i\beta} Y_{\tiny \tilde{R}_2}^{*j\alpha} }{2M_{\tiny \tilde{R}_2}^2}$  \\[3ex]
			\hline
		%%%%%%%%%%%%%%%%%%%%%%%%%%%%%%%%%%
		\multirow{4}{*}{0} & \multirow{4}{*}{$S_1 (\bar{\bf 3},{\bf 1},1/3)$} & \multirow{4}{*}{$+\,\overline{Q^c}\, i \tau_2\, Y_{\tiny S_1} L\; S_1 $} & \multirow{4}{*}{$\dsp\frac{[\mathcal{C}_{lq}^{(1)}]^{ij\alpha \beta}}{\Lambda^2}= -\dsp\frac{[\mathcal{C}_{lq}^{(3)}]^{ij\alpha \beta}}{\Lambda^2}= - \frac{Y_{\tiny S_1}^{j\beta} Y_{\tiny S_1}^{*i\alpha} }{4M_{\tiny S_1}^2}$} &\multirow{4}{*}{ $[N_{LL}^{V,d}]^{ij\alpha\beta}= \dsp \frac{Y_{\tiny {S}_1}^{j\beta} Y_{\tiny {S}_1}^{*i\alpha} }{2M_{\tiny {S}_1}^2}$} \\[7ex]
		&& $+\,\overline{u^c_R}  \,\tilde{Y}_{\tiny S_1}\,  S_1\, e_R$ & $\dsp\frac{[\mathcal{C}_{eu}]^{ij\alpha \beta}}{\Lambda^2}= - \frac{\tilde{Y}_{\tiny S_1}^{j\beta} \tilde{Y}_{\tiny S_1}^{*i\alpha} }{2M_{\tiny S_1}^2} $ & $[N_{RR}^{V,d}]^{ij\alpha\beta}= \dsp \frac{Z_{\tiny {S}_1}^{j\beta} Z_{\tiny {S}_1}^{*i\alpha} }{2M_{\tiny {S}_1}^2}$ \\[2ex]
		&& $+\,\overline{d^c_R}  \,Z_{\tiny S_1}\,  S_1\, \nu_R$  & $\dsp\frac{[\mathcal{C}_{lequ}^{(1)}]^{ij\alpha \beta}}{\Lambda^2}= -\dsp\frac{4[\mathcal{C}_{lequ}^{(3)}]^{ij\alpha \beta}}{\Lambda^2}= - \frac{Y_{\tiny S_1}^{j\beta} \tilde{Y}_{\tiny S_1}^{*i\alpha} }{2M_{\tiny S_1}^2}$ & $[N_{LL}^{S,d}]^{ij\alpha\beta}=-4[N_{LL}^{T,d}]^{ij\alpha\beta}= \dsp \frac{Y_{\tiny {S}_1}^{j\beta} Z_{\tiny {S}_1}^{*i\alpha}  }{2M_{\tiny {S}_1}^2}$  \\[2ex]
		&&  & & $[N_{RR}^{S,d}]^{ij\alpha\beta}=-4[N_{RR}^{T,d}]^{ij\alpha\beta}= \dsp \frac{Z_{\tiny {S}_1}^{j\beta}Y_{\tiny {S}_1}^{*i\alpha}  }{2M_{\tiny {S}_1}^2}$  \\[3ex]
		\hline  \noalign{\vskip 2pt}
		%%%%%%%%%%%%%%%%%%%%%%%%%%%%%%%%%%
		1 & $U_3^\mu ({\bf 3},{\bf 3},2/3)$ & $+\,\overline{Q}\, \gamma^\mu \tau^a\, Y_{\tiny U_1} L\; U_{1\mu}^a$  & $\dsp\frac{[\mathcal{C}_{lq}^{(1)}]^{ij\alpha \beta}}{\Lambda^2}= -\dsp\frac{3[\mathcal{C}_{lq}^{(3)}]^{ij\alpha \beta}}{\Lambda^2}= - \frac{3Y_{\tiny U_3}^{i \beta} Y_{\tiny U_3}^{*j\alpha} }{2M_{\tiny U_3}^2}$  & $[N_{LL}^{V,d}]^{ij\alpha \beta}= - \dsp\frac{2Y_{\tiny U_3}^{i \beta} Y_{\tiny U_3}^{*j\alpha} }{M_{\tiny U_3}^2}$ \\[4ex]
		\hline  %\noalign{\vskip 2pt}
		%%%%%%%%%%%%%%%%%%%%%%%%%%%%%%%%%%
		 \multirow{3}{*}{1} & \multirow{3}{*}{$V_2^\mu (\bar{\bf 3},{\bf 2},5/6)$} & \multirow{3}{*}{$+\,\overline{d^c_R}\, \gamma^\mu Y_{\tiny V_2} V_{2\mu}^T i \tau_2\, L\; $}  & \multirow{3}{*}{$\dsp\frac{[\mathcal{C}_{ld}]^{ij\alpha \beta}}{\Lambda^2}=  \dsp\frac{Y_{\tiny {V}_2}^{i\beta} Y_{\tiny V_2}^{*j\alpha} }{M_{\tiny {V}_2}^2}$} & \multirow{3}{*}{$[N_{RL}^{V,d}]^{ij\alpha \beta}=  \dsp\frac{Y_{\tiny {V}_2}^{j\beta} Y_{\tiny V_2}^{*i\alpha} }{M_{\tiny {V}_2}^2}$} \\[6ex]
		& & $+\,\overline{Q^c_L}\, \gamma^\mu \tilde{Y}_{\tiny V_2}  i \tau_2\, V_{2\mu} e_R\; $ & $\dsp\frac{[\mathcal{C}_{eq}]^{ij\alpha \beta}}{\Lambda^2}=  \frac{\tilde{Y}_{\tiny V_2}^{i\beta} \tilde{Y}_{\tiny V_2}^{*j\alpha} }{M_{\tiny V_2}^2}$ & \\[4ex]
		&& & $\dsp\frac{[\mathcal{C}_{ledq}]^{ij\alpha \beta}}{\Lambda^2}= - \frac{ \tilde{Y}_{\tiny V_2}^{i\beta} Y_{\tiny V_2}^{*j\alpha} }{M_{\tiny V_2}^2}$ & \\[4ex]
		\hline \noalign{\vskip 1pt}
		%%%%%%%%%%%%%%%%%%%%%%%%%%%%%%%%%
		1 & $\bar U_1^\mu ({\bf 3},{\bf 1},-1/3)$ & $+\,\overline{d}_R  \,Z_{\tiny \bar{U}_1}\, \gamma^\mu \bar{U}_{1\mu}\,  \nu_R$ &  %Null
		-- & $[N_{RR}^{V,d}]^{ij\alpha \beta}= -\dsp \frac{Z_{\tiny \bar{U}_1}^{i\beta} Z_{\tiny \bar{U}_1}^{*j\alpha} }{M_{\tiny \bar{U}_1}^2}$ \\ \noalign{\vskip 1pt}
		\hline \hline
	\end{tabular}}
	\caption{Spin, $\mathcal{G}_{\rm SM}=SU(3)_C \otimes SU(2)_L \otimes U(1)_Y$ quantum numbers, and the possible interaction terms (in the Lagrangian) for all scalar and vector leptoquarks mediating $d_j \to d_i \nu_\alpha \bar\nu_\beta $ transitions at tree level involving SM neutrinos as well as the gauge singlet RHN $\nu_R (1,1,0)$. The third and fourth columns list the operators matched at the high scale $\mu=M_{\tiny \rm LQ}$, in the SMEFT basis and the $\nu$-WET basis, respectively, generated by integrating the correspondent mediator.}
	\label{table:LQs}
\end{table}
%%%%%%%%%%%%%%%%%%%%%%%%%%%%%%%%%%%%%%%%%%%%%

\subsection{Leptoquarks}
\label{sec:LQs}

Leptoquarks couple to a lepton and a quark at tree level, and thus at low energies induce interactions between two leptons and two quarks, and/or four leptons(quarks), while the latter is in most cases either stringently suppressed or forbidden in the SM. Leptoquarks can be scalar or vector particles. The analysis with scalar leptoquarks can be done in a model-independent way: however, the phenomenology of vector leptoquarks is much more sensitive to the ultra-violet (UV) completeness of a particular model. The particle content of the full UV theory can in principle affect the low-energy phenomena substantially, hence the results on vector leptoquarks may not be robust. Below are a few examples of NP models with a broad spectrum of new particles, including vector leptoquarks, that can affect the $B\to K^{(*)}\nu \bar{\nu}$ rate via loop effects~\cite{Crivellin:2018yvo,Fuentes-Martin:2020hvc,Cornella:2021sby}.

Incorporating a light SM gauge singlet RHN $\nu_R (1,1,0)$, we list all possible leptoquark interactions contributing to the $d_j \to d_i \nu_\alpha \bar\nu_\beta $ process in Table~\ref{table:LQs}. Here, $Q$ ($L$) denotes the left-handed SM quark (lepton) doublets, while $u_R$ ($d_R$) and $\ell_R$ are the right-handed up-type (down-type) quark and lepton singlets, respectively. The notation $f^c\equiv \mathcal{C}\bar f^{\, T}$ indicates the charge-conjugated field of the fermion $f$.
Here, $Y_{\tiny\rm LQ}$, $\tilde{Y}_{\tiny\rm LQ}$,  and $Z_{\tiny \rm LQ}$ are completely arbitrary Yukawa matrices in flavour space and $\tau_k,~k\in \{1,2,3\}$ are the Pauli matrices. The transformation from the fermion interaction eigenstates to mass eigenstates is simply given by $u_L \to V^\dagger u_L$ where $V$ is the quark Cabibbo-Kobayashi-Maskawa (CKM) matrix \cite{Cabibbo:1963yz,Kobayashi:1973fv}. We have neglected the unitary matrix in the neutrino sector. The generated Wilson coefficients are also shown in the third and fourth columns of Table~\ref{table:LQs} with matching at the high scale $\mu=M_{\tiny \rm LQ}$ in the SMEFT basis and the $\nu$-WET basis, respectively, which are defined in Appendix~\ref{sec:SMEFT}. The evolution using the renormalization group, to the much lower scales at which hadronic decays take place, is also discussed in Appendix~\ref{sec:SMEFT}.

 In the subsequent subsections, we explore each case in which the minimal non-zero set of couplings are chosen that can give a correlation between $b \to s \nu \bar{\nu}$ transition rates and the NC and CC observables in which certain tensions are observed by several experimental collaborations. This immediately eliminates the couplings to first generation quarks and leptons, which are in any case stringently constrained by low energy processes involving lepton and kaon physics~\cite{Davidson:1993qk,Dorsner:2016wpm,Mandal:2019gff,Crivellin:2021egp}. For simplicity, we assume all couplings to be real. For the scalar leptoquarks, notable constraints arise from one loop induced $Z$, $W$ boson decays~\cite{Arnan:2019olv,Falkowski:2019hvp,Crivellin:2020mjs} and $B_s$-mixing data~\cite{DiLuzio:2019jyq} while accommodating one and/or both types of anomaly~\cite{Bordone:2020lnb,Crivellin:2019dwb,Calibbi:2015kma,Carvunis:2021dss}. In the context of $R$-parity violating NP models, which resemble the leptoquark scenarios, the importance of such constraints has been analyzed here~\cite{Altmannshofer:2017poe,Altmannshofer:2020axr,Dev:2021zty,Deshpande:2012rr,Deshpande:2016yrv}. We consider these constraints and the anomalies in the subsequent sections. 

Leptoquarks have been extensively searched for at colliders; so far, the  absence of any evidence of a signal imposes limits on the mass range. The limits are stronger for leptoquarks decaying to first generations of fermions~\cite{Aad:2020iuy} compared to those for third generation final states~\cite{Aad:2021rrh,CMS:2020wzx}. For the remainder of the analysis, we choose a benchmark case of a $2\,$TeV mass for all leptoquarks, which will be probed in upcoming analyses at the LHC.

\subsubsection{$\mathbf{S_3}(\bar{\bf 3},{\bf 3},1/3)$}

The triplet scalar leptoquark $S_3(\bar{\bf 3},{\bf 3},1/3)$ contributes to both $b\to s \mu \mu$ and $b \to c \tau \bar \nu$ transitions via tree level interactions. From Eq.~\eqref{eq:WCs_bsll}, we have the NP contribution to $b\to s \mu \mu$: 
\bea
C_9^{\rm NP}= - C_{10}^{\rm NP}=  \frac{ v^2}{M_{S_3}^2} \frac{\pi}{\alpha_\text{EM} V_{tb}V_{ts}^*} Y_{\tiny S_3}^{22} Y_{\tiny S_3}^{*32}\,.
\eea
In the case of $b\to c \tau \bar{\nu}_\beta$, from Eq.~\eqref{eq:btoc_coeffs} we obtain
\bea
\mathcal{P}_{LL}^{V 3\beta} = - \frac{v^2}{4M_{\tiny{S_3}}^2} Y_{\tiny S_3}^{3\beta}\left[Y_{\tiny S_3}^{*33}+\frac{1}{V_{cb}}\left(V_{cs} Y_{\tiny S_3}^{*23} + V_{cd} Y_{\tiny S_3}^{*13} \right)\right] \,.
\eea
For $b \to s \nu^\alpha \bar{\nu}^\beta$ the Wilson coefficient quoted in Eq.~\eqref{eq:Heff} is found to be
\bea
[{C}_{LL}^V]^{\alpha\beta}=\,\frac{v^2}{ M_{\tiny S_3}^2}\frac{\pi}{\alpha_\text{EM} V_{tb}V_{ts}^*}Y_{\tiny S_3}^{ 3\beta}Y_{\tiny S_3}^{*2\alpha}\,.
\eea

The minimal set of non-zero couplings required in order to generate contributions to both CC and NC anomalies is $Y_{\tiny S_3}^{22},\,Y_{\tiny S_3}^{32},\,Y_{\tiny S_3}^{33}$ and $Y_{\tiny S_3}^{23}$. The desired values for the couplings $Y_{\tiny S_3}^{33}$ and $Y_{\tiny S_3}^{23}$ to resolve the tension in the CC channels are ruled out by the $B_s-\bar B_s$ mixing bound at a significance level greater than $2\sigma$ (95\,\% C.L.). However, in order to explain the $b\to s \mu \mu$  tensions via $C_9^{\rm NP}=-C_{10}^{\rm NP}=-0.41^{+0.07}_{-0.07}$~\cite{Altmannshofer:2021qrr}, we obtain the following ranges for the $S_3$ coupling constants for $m_{S_3}=2\,$TeV:
\bea
Y_{\tiny S_3}^{32}Y_{\tiny S_3}^{22} = 0.0028\pm 0.0005\,,\qquad |Y_{\tiny S_3}^{32}|\le 1.33\,.
\eea
Here the bound on $Y_{\tiny S_3}^{32}$ arises from the $Z\to \nu \bar{\nu}$ constraint, which is stronger than the $Z\to \mu \mu$ bound in this case. This entire parameter space is compatible with the current data on $R_K^\nu$, however, only a  2\% enhancement in $B \to K \nu \bar{\nu}$ can be achieved. If an improved measurement results in a larger enhancement in $B \to K \nu \bar{\nu}$, the only option would be  to turn on other elements in the leptoquark Yukawa matrix  in order to be compatible with the signal. As a benchmark case, we explore the situation with third generation lepton couplings and find the following best fit values:
\bea
Y_{\tiny S_3}^{22}= 0.025 \pm 0.024\,,\quad  Y_{\tiny S_3}^{23}= 1.0\pm 0.83\,,\quad Y_{\tiny S_3}^{32} = 0.109\pm 0.085\,,\quad Y_{\tiny S_3}^{33}= -0.041\pm 0.024\,
\eea 
resulting in $R_K^\nu= 2.4\pm 3.6$, which is 
consistent with the experimental central value in Eq.~\eqref{eq:RKnu}. Thus, although we see hardly any impact on CC anomalies with the couplings $Y_{\tiny S_3}^{23}$ and $Y_{\tiny S_3}^{33}$, their contribution to $R_K^\nu$ can be important.

%%%%%%%%%%%%%%%%%%%%%%%%%%%%%%%%%%%%%
\subsubsection{$\mathbf{\tilde{R}_2}(\bar{\bf 3},{\bf 2},1/6)$}

Next, we consider the doublet scalar leptoquark $\tilde{R}_2(\bar{\bf 3},{\bf 2},1/6)$, which also contributes to both $b\to s \mu \mu$ and $b \to c \tau \bar \nu_\beta $ transitions, where the latter proceeds only via light RHN interactions.	From Eqs.~\eqref{eq:WCs_bsll} and \eqref{eq:Heffb2c}, we have the following contributions:
\begin{align}
C_{9}^{\prime\, 22} = - C_{10}^{\prime\, 22} & = -\dsp\frac{v^2}{2M_{\tiny \tilde{R}_2}^2}\frac{\pi}{\alpha_\text{EM} V_{tb}V^*_{ts}}Y_{\tiny \tilde{R}_2}^{22}Y_{\tiny \tilde{R}_2}^{*32}\,,\\
\mathcal{P}_{RR}^{S\,3\beta} (M_{\tiny \tilde{R}_2}) =4  \mathcal{P}_{RR}^{T\,3\beta} (M_{\tiny \tilde{R}_2}) & = \dsp\frac{v^2}{4 M_{\tiny \tilde{R}_2}^2}  Y_{\tiny \tilde{R}_2}^{*3 3} \left[Z_{\tiny \tilde{R}_2}^{ 3\beta} + \frac{V_{cs}}{V_{cb}} Z_{\tiny \tilde{R}_2}^{ 2\beta} + \frac{V_{cd}}{V_{cb}} Z_{\tiny \tilde{R}_2}^{ 1\beta}  \right]\,.
\end{align}
For $b \to s \nu \bar{\nu}$ we have four different operators as shown in Table \ref{table:LQs}, related to the Wilson coefficients in Eq.~\eqref{eq:Heff} via 
\begin{align}
\label{eq:NEFT_facs}
[C_{AB}^X]^{\alpha \beta} = 2 v^2 \frac{\pi}{\alpha_\text{EM} V_{tb}V^*_{ts}} [N_{AB}^{X,d}]^{23\alpha \beta}\,.
\end{align}

\begin{figure}[h!]
	\centering
	\includegraphics[scale=0.8]{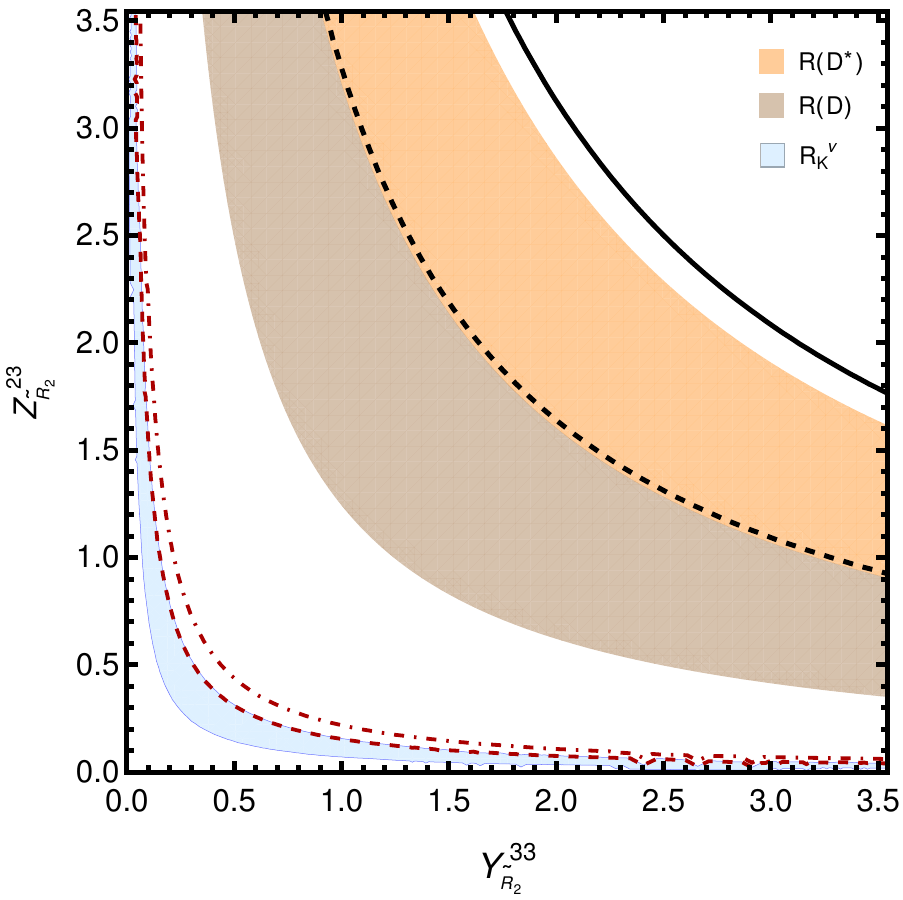}
	\caption{The regions in the $Y_{\tiny \tilde{R_2}}^{33}-Z_{\tiny \tilde{R_2}}^{23}$ plane show a lack of compatibility with the desired range for $R(D^{*})$ anomalies and the current limit from the $B^+ \to K^+ \nu \bar{\nu}$ mode. The red dashed and dot-dashed curves denote cases in which $R_{K^*}^\nu$ equals 2 or 3 respectively, whereas the region below the black dotted(solid) curve is allowed from the $\mathcal{B}(B_c \to \tau \bar\nu)\lesssim10\%(30\%)$ limit. }
	\label{fig:R2t}
\end{figure}

Although $\tilde{R_2}$ affects the NC transitions, the contribution
$C_9^\prime=-C_{10}^\prime$ can merely reduce the tension in $R_K$ and gives $R_{K^*}>1$~\cite{Angelescu:2021lln} in the region $[1,6]\,\gev^2$, which disagrees with the data at greater than 99\% C.L. and hence the NC anomalies cannot be accommodated in this scenario. The minimal non-zero couplings required for the CC anomalies are $Y_{\tiny \tilde{R}_2}^{33}$ and $Z_{\tiny \tilde{R}_2}^{23}$ and the corresponding parameter space is shown in Fig.~\ref{fig:R2t}. It can be seen that the $1\sigma$ compatible region for the CC anomalies is excluded by the $\mathcal{B}(B\to K \bar\nu\nu) $ data. We note that such a contribution arises from a scalar structure ($[C_{LL}^S]^{33}$ in Eq.~\eqref{eq:Heff}) that is generated via the $\tilde{R_2}$ leptoquark interaction terms involving RHNs. This scalar operator does not interfere with the SM terms and hence it is not possible to cancel this large contribution by some other effects.  If we include $R_K^\nu$ in the $\chi^2$-fit no enhancement in $R(D^{(*)})$ is seen and hence the discrepancy cannot be explained. To summarize, $\tilde{R_2}$ can produce the $R_K^\nu$ future scenario, however the $\tilde{R_2}$ parameter space cannot resolve any of the tensions observed in the $b\to s \mu \mu$ and $b \to c \tau \bar{\nu}$ channels.

\subsubsection{$\mathbf{S_1}(\bar{\bf 3},{\bf 1},1/3)$}

This particular scalar leptoquark does not contribute to $b \to s \mu \mu$ transition at tree level and for the CC transition $b \to c \tau \bar\nu_\beta$ we have the following  set of operators involving both the SM neutrinos and RHNs:
\begin{align}
\mathcal{P}_{LL}^{V 3\beta} &= \frac{v^2}{4M_{\tiny{S_1}}^2} Y_{\tiny S_1}^{3\beta}\left[Y_{\tiny S_1}^{*33}+\frac{1}{V_{cb}}\left(V_{cs} Y_{\tiny S_1}^{*23} + V_{cd} Y_{\tiny S_1}^{*13} \right)\right] \,, \\
\mathcal{P}_{LL}^{S\,3\beta} (M_{\tiny S_1}) &=-4  \mathcal{P}_{LL}^{T\,3 \beta} (M_{\tiny S_1})  = - \dsp\frac{v^2}{4 M_{\tiny S_1}^2} \frac{1}{V_{cb}} Y_{\tiny S_1}^{3 \beta} \tilde{Y}_{\tiny S_1}^{ *23}\,, 
\end{align}
\begin{align}
\mathcal{P}_{RR}^{V\,3 \beta} &= - \dsp\frac{v^2}{4 M_{\tiny S_1}^2} \frac{1}{V_{cb}}  Z_{\tiny S_1}^{3 \beta}\tilde{Y}_{\tiny S_1}^{ *23} \,,  \\
\mathcal{P}_{RR}^{S\,3\beta} (M_{\tiny S_1}) &=-4  \mathcal{P}_{RR}^{T\,3 \beta} (M_{\tiny S_1})  =  \dsp\frac{v^2}{4 M_{\tiny S_1}^2}  Z_{\tiny S_1}^{*3 \beta}\left[Y_{\tiny S_1}^{*33}+\frac{1}{V_{cb}}\left(V_{cs} Y_{\tiny S_1}^{*23} + V_{cd} Y_{\tiny S_1}^{*13} \right)\right]\,.
\end{align}
For $b \to s \nu \bar{\nu}$, we have four different operators (given in Table~\ref{table:LQs}) and they follow the relation quoted in Eq.~\eqref{eq:NEFT_facs}. It is evident from the above set of equations that there exist several possible choices of NP couplings that relate the $b\to c \tau \bar{\nu}$ channel with $b \to s \nu \bar{\nu}$. Below we study these scenarios case by case.

{\it Scenario $S_1$-I:} We start with the minimal set of purely left-handed couplings for both the quark and lepton sectors i.e., $Y_{\tiny S_1}^{33}$ and $Y_{\tiny S_1}^{23}$. In this case $Z\to \tau \tau$ and $\Delta m_s$ data restrict large values required in order to explain the $R(D^{*})$ discrepancies. The best fit values with $\pm 1\sigma$ uncertainties for $M_{S_1}=2\,$TeV including $R(D^{*})$, $F_L^{D^*}$, $\Delta m_s$, $Z\to \tau \tau$, and $Z\to\nu \nu$ data read
\bea
Y_{\tiny S_1}^{ 33}= 1.72\pm 0.67\,,\quad Y_{\tiny S_1}^{ 23}= 0.07\pm 0.03\,,
\eea
showing a 4\% enhancement in $R(D^{*})$ values that cannot reduce the tension to the $1\sigma$ level for these observables. There exists a sign flipped minima at which both the coupling values are negative. This parameter space is compatible with the current $R_K^\nu$ data (Eq.~\eqref{eq:RKnu}) and reproduces the range of $R_K^\nu= 2.46\pm 1.22$.

{\it Scenario $S_1$-II:}  Next, we look to accommodate the CC anomalies via right-handed couplings but restricting ourselves only to SM neutrinos. In this case the minimal set of non-zero couplings required is $Y_{\tiny S_1}^{33}$ and $\tilde{Y}_{\tiny S_1}^{23}$. A $\chi^2$ fit including $R(D^{(*)}),\,\Delta m_s$ and $Z$-decays data can explain both $R(D)$ and $R(D^*)$ within $\pm1\sigma$ uncertainties, however, this scenario does not contribute to $\mathcal{B}(B\to K \bar\nu\nu)$.
Allowing one  more SM neutrino coupling namely $Y_{\tiny S_1}^{23}$, we generate a contribution to $\mathcal{B}(B\to K \bar\nu\nu)$ and the parameter space can explain both $R(D)$, $R(D^*)$ and $R_K^\nu$ data within their $1\sigma$ uncertainties. The best fit values with $\pm 1\sigma$ uncertainties  in this case are
\bea
Y_{\tiny S_1}^{ 33}= 1.05\pm 0.92\,,\quad Y_{\tiny S_1}^{ 23}= 0.10\pm 0.06\,,\quad \tilde{Y}_{\tiny S_1}^{ 23}= -0.69\pm 0.43\,,
\eea
and we obtain the contribution to $R_K^\nu= 2.35\pm 1.97$.

\begin{figure}[h !]
	\centering
	\includegraphics[scale=0.8]{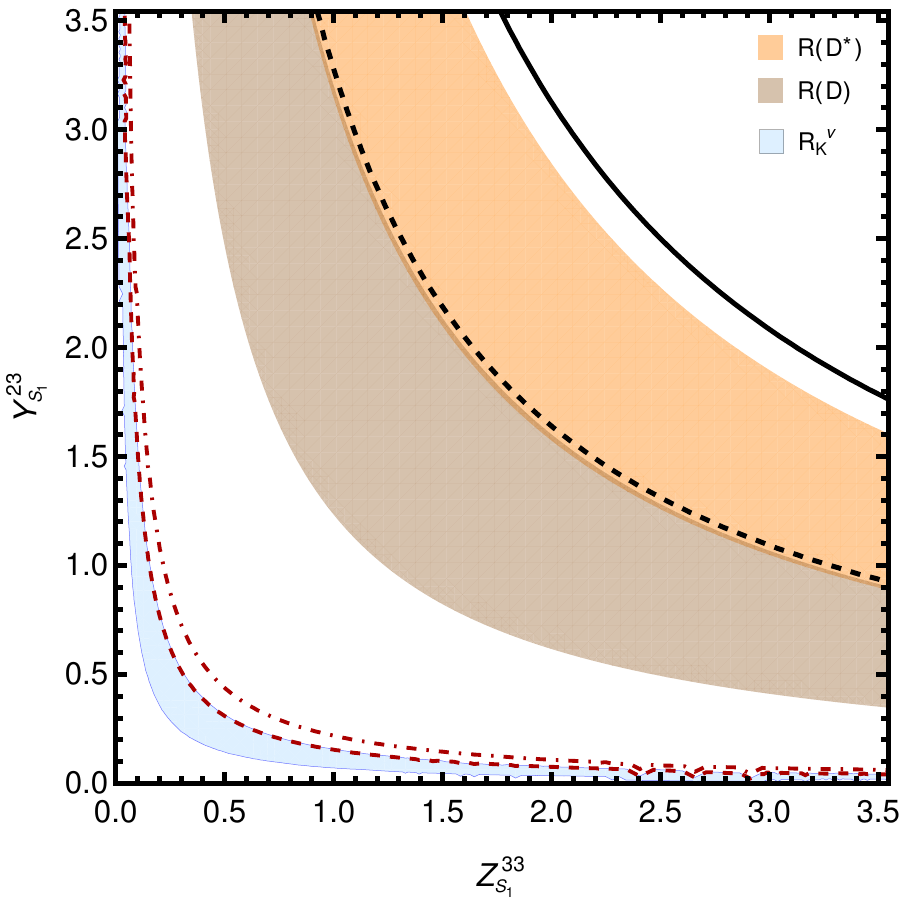}
	\caption{The regions corresponding to the scenario {\it $S_1$-III} in the $Y_{\tiny S_1}^{33}-Z_{\tiny S_1}^{23}$ plane show a lack of compatibility  with the desired $1\sigma$ range for $R(D^{*})$ anomalies and the limit from the $B^+ \to K^+ \nu \bar{\nu}$ mode. The red dashed and dot-dashed curves denote cases in which $R_{K^*}^\nu$ equals 2 or 3 respectively, whereas the region below the black dotted(solid) curve is allowed from the $\mathcal{B}(B_c \to \tau \bar\nu)\lesssim10\%(30\%)$ limit. }
	\label{fig:S1}
\end{figure}

{\it Scenario $S_1$-III:} Allowing for both SM neutrino and RHN couplings, the situation in the minimal set of non-zero coupling (required for CC anomalies) plane  $Y_{\tiny S_1}^{23}$ and $Z_{\tiny S_1}^{33}$ is shown in Fig.~\ref{fig:S1}. Similar to the previous case of $\tilde{R}_2$, we find the $1\sigma$ compatible regions for CC anomalies are not allowed by the $R_K^\nu$ data. Hence this scenario does not provide a viable solution to the CC discrepancies. Going one step towards a next to minimal scenario with non-zero $\tilde{Y}_{\tiny S_1}^{33}$ (which does not contribute to $\mathcal{B}(B\to K \bar\nu\nu)$ but to $R(D^{(*)})$) does not improve the situation.

\subsubsection{ $\mathbf{U_3^{\mu}}(\bar{\bf 3},{\bf 3},2/3)$ }
Now moving to the vector leptoquarks, $\mathbf{U_{3\mu}}(\bar{\bf 3},{\bf 3},2/3)$ contributes to both the NC and CC modes. Using Eq.~\eqref{eq:WCs_bsll} we find
\bea
C_9^{\rm NP}= - C_{10}^{\rm NP}=  -\frac{ v^2}{M_{U_3}^2} \frac{\pi}{\alpha_\text{EM} V_{tb}V_{ts}^*} Y_{\tiny U_3}^{22}Y_{\tiny U_3}^{*32}\,.
\eea
The non-vanishing Wilson coefficient in Eq.~\eqref{eq:Heffb2c} is
\bea
\label{eq:PVU3}
\mathcal{P}_{LL}^{V,3\beta} =- \dsp\frac{v^2}{2 M_{\tiny U_3}^2}  Y_{\tiny U_3}^{*3 3} \left[Y_{\tiny U_3}^{ 3\beta} + \frac{V_{cs}}{V_{cb}} Y_{\tiny U_3}^{ 2\beta} + \frac{V_{cd}}{V_{cb}} Y_{\tiny U_3}^{ 1\beta}  \right]\,,
\eea
and in Eq.~\eqref{eq:Heff} is found to be
\bea
[{C}_{LL}^V]^{\alpha\beta}=\,-\frac{4v^2}{ M_{\tiny U_3}^2}\frac{\pi}{\alpha_\text{EM}V_{tb}V_{ts}^*}Y_{\tiny U_3}^{ 2\beta}Y_{\tiny U_3}^{*3\alpha}\,.
\eea

As mentioned above, in the vector leptoquark case we only examine the constraints arising from tree level processes, as the loop induced decays might be affected by the field content of the UV completion of the vector leptoquark under consideration.
It is clear from the above expressions that the minimal set of non-zero couplings required for the CC and NC anomalies are $Y_{\tiny U_3}^{33},\,Y_{\tiny U_3}^{32},\,Y_{\tiny U_3}^{22}$ and $Y_{\tiny U_3}^{23}$. Note that for $R(D^{*})$ we need $\mathcal{P}_{LL}^{V,33}>0$, however, due to the presence of an overall negative sign in the Wilson coefficient (in Eq.~\eqref{eq:PVU3}), a non-zero $Y_{\tiny U_3}^{23}$ is necessary that can compensate for the reduction arising from the ${(Y_{\tiny U_3}^{33})}^2$ term.  The best fit to $b\to c \tau \bar{\nu}$ and $b\to s \mu \mu$ data overshoots the bound from $R_k^\nu$ by 2 orders of magnitude, while the inclusion of current $R_k^\nu$  data in the fit can explain $b\to s \mu \mu$ anomalies within their $1\sigma$ uncertainties. However, we only see at most a  1\% enhancement for $R(D^{(*)})$, and hence cannot explain the CC anomalies within their $\pm 1\sigma$ uncertainties. Such a region is compatible with the future anticipated sensitivity for $R_K^\nu$.

\subsubsection{ $\mathbf{V_2^{\mu}}(\bar{\bf 3},{\bf 2},5/6)$} 
This vector leptoquark also contributes to both the NC and CC modes. For $b \to s \mu \mu$ (Eq.~\eqref{eq:WCs_bsll}) we obtain
\begin{align}
&C_9^{\rm NP}=  C_{10}^{\rm NP}= - \frac{ v^2}{M_{V_2}^2} \frac{\pi}{\alpha_\text{EM}V_{tb}V_{ts}^*} \tilde{Y}_{\tiny V_2}^{*32} \tilde{Y}_{\tiny V_2}^{22}\,, \\
&C_9^\prime= - C_{10}^\prime=  -\frac{ v^2}{M_{V_2}^2} \frac{\pi}{\alpha_\text{EM} V_{tb}V_{ts}^*} Y_{\tiny V_2}^{*32}Y_{\tiny V_2}^{22}\,, \\
&C_S= -C_P=  \frac{ 2 v^2}{M_{V_2}^2} \frac{\pi}{\alpha_\text{EM} V_{tb}V_{ts}^*} \tilde{Y}_{\tiny V_2}^{*32}Y_{\tiny V_2}^{22}\,, \\
&C_S^\prime= C_P^\prime=  \frac{ 2 v^2}{M_{V_2}^2} \frac{\pi}{\alpha_\text{EM} V_{tb}V_{ts}^*} Y_{\tiny V_2}^{*32}\tilde{Y}_{\tiny V_2}^{22}\,.
\end{align}
The tree level matching to Wilson coefficients in Eq.~\eqref{eq:Heffb2c} yields
\bea
\mathcal{P}_{RL}^{S,3\beta} = \dsp\frac{v^2}{M_{\tiny V_2}^2}  Y_{\tiny V_2}^{3 3} \left[\tilde{Y}_{\tiny V_2}^{ *3\beta} + \frac{V_{cs}}{V_{cb}} \tilde{Y}_{\tiny V_2}^{ *2\beta} + \frac{V_{cd}}{V_{cb}} \tilde{Y}_{\tiny V_2}^{* 1\beta}  \right]\,,
\eea
and for the operators in Eq.~\eqref{eq:Heff} we find
\bea
[{C}_{RL}^V]^{\alpha\beta}=\,\frac{2 v^2}{ M_{\tiny V_2}^2}\frac{\pi}{\alpha_\text{EM} V_{tb}V_{ts}^*}Y_{\tiny V_2}^{ 3\beta}Y_{\tiny V_2}^{*2\alpha}\,.
\eea

Here, neither of the scenarios $C_9^{\rm NP}=C_{10}^{\rm NP}$ and $C_9^\prime=-C_{10}^\prime$ explains the $b\to s \mu \mu$ data (barely preferred in the global fits~\cite{Alguero:2021anc}), thus an explanation of the NC anomalies can only be obtained to a certain extent~\cite{Altmannshofer:2021qrr} via very small values of $C_S^{(\prime)},\,C_P^{(\prime)}$ i.e. $C_S=-C_P=-0.003^{+0.002}_{-0.002}$ and $C_S^\prime=C_P^\prime=-0.003^{+0.002}_{-0.002}$. Therefore we choose the minimal set of non-zero couplings required for CC and NC anomalies as follows: $Y_{\tiny V_2}^{23,33},\,\tilde{Y}_{\tiny V_2}^{22,32}$. Interestingly, in the case of CC anomalies, the Wilson coefficient $\mathcal{P}_{RL}^{S,33}$ can only accommodate the discrepancy in $R(D)$ but not in $R(D^*)$ within their $\pm1\sigma$ uncertainties. This is due to the different dependency of $\mathcal{P}_{RL}^{S,33}$ on the corresponding observables for pseudoscalar $D$ and vector $D^*$ modes as can be seen from Eqs.~\eqref{eq:RDRDSM} and \eqref{eq:RDstRDSM}. Such a parameter space can accommodate $R_K^\nu\simeq2.4$.

\subsubsection{$\mathbf{\bar{U}_1^{\mu}}({\bf 3},{\bf 1},-1/3)$}
It is clear from the interaction term given in Table~\ref{table:LQs} that, among the decays of our interest, this particular vector leptoquark affects $b \to s \nu \bar \nu$ mode only and has no correlation with the modes where anomalies are observed. From Eq.~\eqref{eq:Heff} we obtain
\bea
[{C}_{RR}^V]^{\alpha\beta}=\,-\frac{2v^2}{ M_{\tiny \bar{U}_1}^2}\frac{\pi}{\alpha_\text{EM} V_{tb}V_{ts}^*}Z_{\tiny \bar{U}_1}^{ 2\beta}Z_{\tiny \bar{U}_1}^{*3\alpha}\,.
\eea
Now from Fig.~\ref{fig:var}, we can see that the current and expected $R_K^\nu$ data can easily be explained for a wide range of the Wilson coefficient $[{C}_{RR}^V]^{\alpha\alpha}$ and there are no other constraints on these Wilson coefficients.

\subsection{Generic $Z^\prime_\mu$}
\label{sec:Z'}

In this section we consider a general $Z^\prime$ model with a tree level FCNC transition~\cite{Buras:2012jb}. Since $Z^\prime$ exchange involves a neutral vector boson directly exchanged between quarks and leptons, it can only affect $b\to s \mu \mu$ and $b\to s \nu \bar{\nu}$ transitions not $b\to c \tau \bar\nu$. As we show below, it is possible to account for $b\to s \mu \mu$ data by a suitable choice of flavor changing $Z^\prime$ couplings. However in all cases, the rate for $b\to s \nu \bar{\nu}$ is modified by at most 10\%. 
The basic Lagrangian describing the interactions of the $Z^\prime$ gauge boson
with SM fermions in the flavor basis can be written as~\cite{Buras:2012jb}
\bea
\mathcal{L}\left(Z^{\prime}\right)=\sum_{i, j, \psi_{L}} \Delta_{L}^{i j} \bar{\psi}_{L}^{i} \gamma^{\mu} P_{L} \psi_{L}^{j} Z_{\mu}^{\prime}+\sum_{i, j, \psi_{R}} \Delta_{R}^{i j} \bar{\psi}_{R}^{i} \gamma^{\mu} P_{R} \psi_{R}^{j} Z_{\mu}^{\prime}\,,
\eea
where $\psi$ represents fermions with the same electric charge and $i,\,j$ are the generation indices. The couplings matched with the SMEFT basis lead to
\begin{align}
[\mathcal{C}^{(1)}_{lq}]^{ij\alpha\beta} =&  - \Delta_L^{ij} \Delta_L^{\alpha\beta}\,,&
[\mathcal{C}_{ld}]^{ij\alpha\beta} =&  - \Delta_R^{ij} \Delta_L^{\alpha\beta}\,,& \\
[\mathcal{C}_{eq}]^{ij\alpha\beta} =&  - \Delta_L^{ij} \Delta_R^{\alpha\beta}\,,&
[\mathcal{C}_{ed}]^{ij\alpha\beta} =&  - \Delta_R^{ij} \Delta_R^{\alpha\beta}\,.&
\end{align}

As we intend to explore the correlation with $b\to s$ semileptonic transitions, the relevant couplings are $\Delta_{L,R}^{sb}$, $\Delta_{L,R}^{\ell\ell}$ and $\Delta_{L}^{\nu\nu}= \Delta_{L}^{\ell\ell}$ where the relation at the $Z^\prime$ scale follows from $SU(2)_L$ gauge invariance. One of the strong constraints on $\Delta_{L,R}^{bs}$  arises from  $B_s-\bar{B}_s$ mixing where the contribution to the mass difference $\Delta M_s$ can be written as~\cite{DiLuzio:2019jyq}
\bea
\label{eq:MsexpZp}
\frac{\Delta M_{s}^{\mathrm{SM}+\mathrm{NP}}}{\Delta M_{s}^{\mathrm{SM}}} \approx\left|1+200\left(\frac{5 \mathrm{TeV}}{M_{Z^{\prime}}}\right)^{2}\left[\left(\Delta_L^{sb}\right)^{2}+\left(\Delta_R^{sb}\right)^{2}-9 \Delta_L^{sb} \Delta_R^{sb}\right]\right|\,.
\eea
Here the numerical factors include the renormalization group effects and the factor 9 is the ratio of four-quark matrix elements evaluated at a scale $M_{Z^\prime}=5\,$TeV. We use the results in Ref.~\cite{DiLuzio:2019jyq}, where the weighted average of SM predictions is used to obtain
\begin{equation}
\begin{aligned}
\Delta M_s^\text{average} =&\, (1.04^{+0.04}_{-0.07})\, \Delta M_s^\text{exp}\,. 
\end{aligned}
\end{equation}
It can be easily seen from Eq.~\eqref{eq:MsexpZp} that the couplings $\Delta_{L,R}^{sb}$ are stringently constrained by the $B_s-\bar B_s$ mixing data; whereas the leptonic couplings especially to the second and  third generations can take up to $\mathcal{O}(1)$ values~\cite{Buras:2021btx} for a TeV mass $Z^\prime$, evading the bounds from electroweak precision observables, LEP data, rare lepton decays,  neutrino trident production etc. The limits from the latest direct searches are somewhat less constraining ($m_{Z^\prime}>1.5\,$TeV~\cite{Allanach:2019mfl}) for a minimal $Z^\prime$ setup, which can accommodate the NC anomalies and thus our choice of $M_{Z^\prime}=5\,$TeV lies on the conservative side.

First we consider the case with only left-handed couplings and the minimal choice is $\Delta_{L}^{sb}$ and $\Delta_{L}^{\mu\mu}(=\Delta_{L}^{\nu_\mu\nu_\mu})$. In this scenario, we generate
\bea
C_9^{\rm NP} = -C_{10}^{\rm NP} =  \frac{ v^2}{M_{Z^\prime}^2} \frac{\pi}{\alpha_\text{EM} V_{tb}V_{ts}^*}\, \Delta_{L}^{sb} \Delta_{L}^{\mu\mu} \,.
\eea
Assuming all couplings to be real, we can obtain a good fit to the $b\to s \mu \mu$  tensions via $C_9^{\rm NP}=-C_{10}^{\rm NP}=-0.41^{+0.07}_{-0.07}$~\cite{Altmannshofer:2021qrr} for $\Delta_{L}^{sb}  = \left(8.5 \pm 6.4 \right)\times 10^{-3}$ and $\Delta_{L}^{\mu\mu} =2.00\,\pm\, 0.95$ while being consistent with all data. Now this parameter space is compatible with the current $R_K^\nu$ data, however, at most it can give $R_K^\nu=1.05 \pm 0.03$, which is too close to the SM to be distinguished in the near future.

Next we turn on the right-handed quark coupling $\Delta_{R}^{sb} $ that corresponds to
\bea
C_9^{\rm NP} = -C_{10}^{\rm NP} =  \frac{ v^2}{M_{Z^\prime}^2} \frac{\pi}{\alpha_\text{EM} V_{tb}V_{ts}^*}\, \Delta_{L}^{sb} \Delta_{L}^{\mu\mu} \,, \\
C_9^\prime = - C_{10}^\prime = \frac{ v^2}{M_{Z^\prime}^2} \frac{\pi}{\alpha_\text{EM} V_{tb}V_{ts}^*}\, \Delta_{R}^{sb} \Delta_{L}^{\mu\mu} \,.
\eea
This 2D scenario for the NP Wilson coefficients is also preferred in the global fit to all $b\to s \mu \mu$ data~\cite{Alguero:2021anc} and in accommodating the discrepancies, however the effect on $b\to s \nu \bar\nu$ remains unaltered compared to the previous scenario and predicts $R_K^\nu\approx 1.1$.

\section{Discussion}
\label{sec:dis}

\begin{figure}[h !]
	\centering
	\includegraphics[scale=0.79]{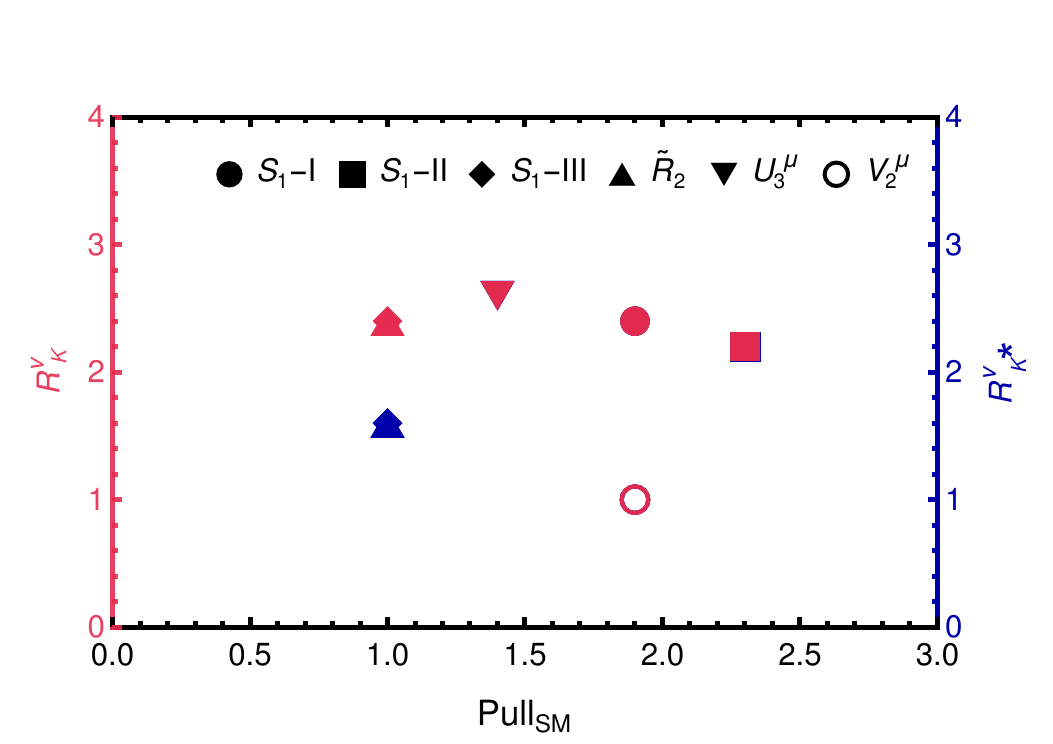} %\hskip 20pt
	\includegraphics[scale=0.79]{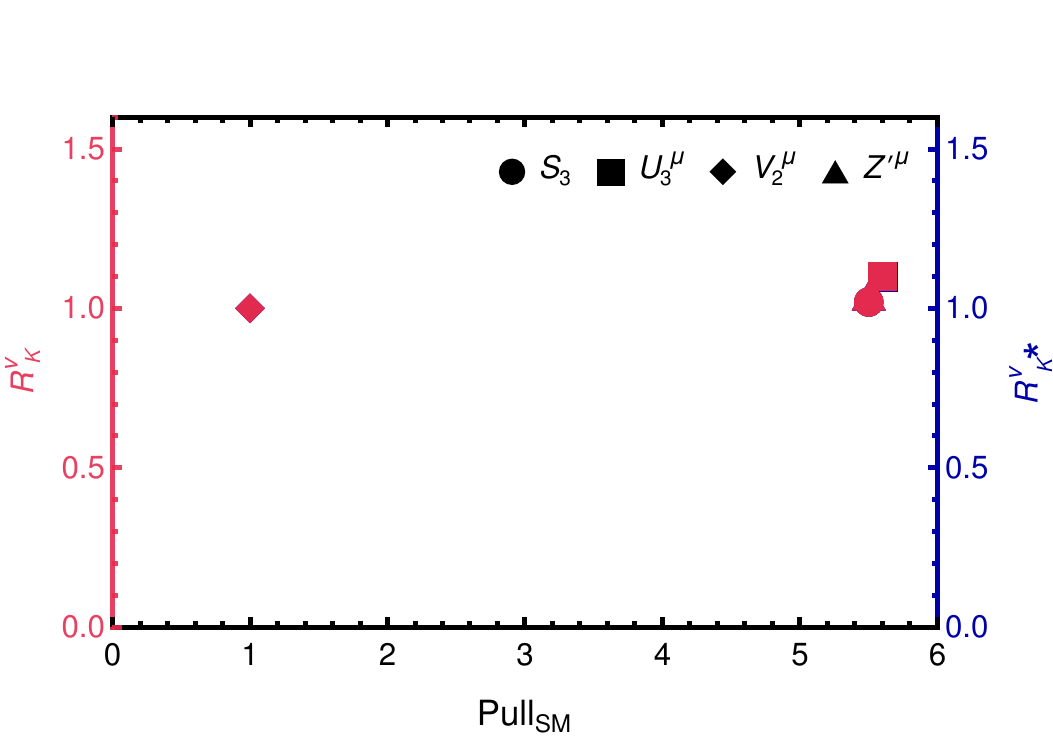} \vskip -10pt
	\includegraphics[scale=0.77]{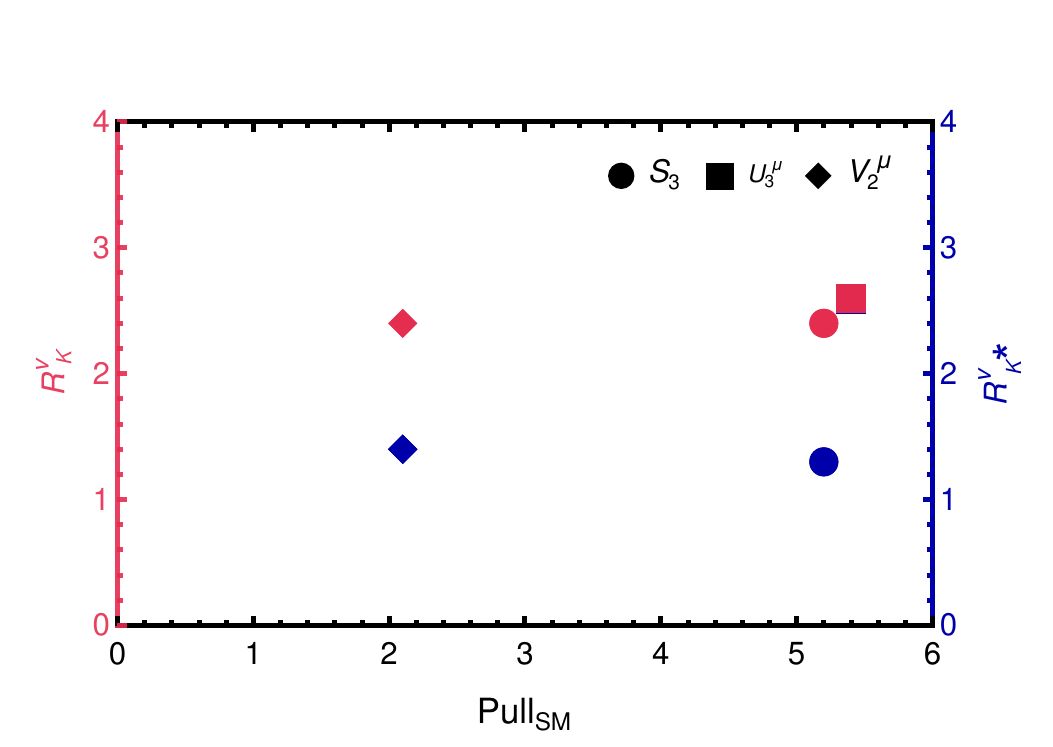}
	\caption{The summary of NP scenarios in the context of ``${\rm Pull_{SM}} $" values for the fits to CC, NC and combined CC and NC discrepancies together with their predictions for $R_{K^{(*)}}^\nu$ are shown in the top-left, top-right and bottom panels, respectively. In each panel, the left (in red) and right $y$-axes (in blue) denote the predictions for $R_K^\nu$ and $R_{K^*}^\nu$, respectively, where in most cases the points overlap. The different NP mediators with the corresponding scenarios discussed in Secs.~\ref{sec:LQs} and \ref{sec:Z'}  are depicted by different markers (as highlighted in the legends), where the masses for all leptoquarks are chosen to be 2\,TeV and for the $Z^\prime$ as 5\,TeV. In the case of the mediators contributing to both the CC and NC anomalies i.e., $S_3$, $U_3^\mu$ and $V_2^\mu$, apart from the combined results (in the lower panel) we have also performed fits separating the two situations with a minimal coupling setup. }
	\label{fig:pull}
\end{figure}

We summarize the findings of the previous sections here. In order to quantify the fit to the CC and/or NC anomalies and the impact on $b\to s \nu \bar\nu$ modes, a quantity ``${\rm Pull_{SM}}$" is defined as~\cite{Capdevila:2018jhy}
\bea
{\rm Pull_{SM}} = \sqrt{2}\, {\rm Erf}^{-1}[1- p(\Delta \chi^2,\Delta n)]\,,
\eea
where for any particular NP scenario with $n$ degrees of freedom (d.o.f), 
$p( \chi_{\rm min}^2, n)= \int_{\chi_{\rm min}}^{\infty} dx\, \chi^2(x,n)$ known as the $p$-value and $\Delta \chi^2=\chi_{\rm SM}^2-\chi_{\rm min}^2$ is assumed to follow a $\chi^2$ distribution with $\Delta n = (n_{\rm SM}-n)$ d.o.f.\,. Here ``${\rm Pull_{SM}} $" represents the comparison of any fitted solution with the SM results in units of $\sigma$. A larger value implies a better explanation of the data. We have seen that the various NP scenarios described in previous sections can either contribute to one or both types of anomalies seen in $B$-decays. We show the interplay of these NP fits of these anomalies and the corresponding effect in $b\to s \nu \bar{\nu}$ modes in Fig.~\ref{fig:pull} where the left $y$-axis (in red) denotes the prediction for $R_K^\nu$ and the right one (in blue) $R_{K^*}^\nu$. The different NP cases are depicted by different markers (as described in the legend in the figure) where the masses for all leptoquarks are chosen to be 2\,TeV and the $Z^\prime$ mass is taken as 5\,TeV as discussed in the previous sections.  We have separately shown the results for CC, NC and combined CC and NC setups in the three panels. 

From the top-left panel, we find the two scenarios for the $S_1$ leptoquark ($S_1$-I and II) that have a ``${\rm Pull_{SM}} $" close to $2\sigma$ and $2.5\sigma$, respectively, and also provide $R_{K^{(*)}}^\nu \approx 2.4$ whereas a SM neutrino together with a RHN case for $S_1$ ($S_1$-III), $\tilde{R}_2$, $V_2^\mu$ and $U_3^\mu$ are not preferred solutions for these anomalies. The top-right panel shows fit results for the NC anomalies where the three scenarios namely $S_3$, $U_3$ and $Z^\prime$ show quite a bit larger ``${\rm Pull_{SM}} $" values (greater than $5\sigma$); however, in all these scenarios we barely see any enhancement in $R_{K^{(*)}}^\nu$. Finally, the lower panel shows results for the combined fit when these NP mediators contribute to both the two types of anomalies. Note that while comparing this lower panel with the top-right panel for $S_3$ and $U_3^\mu$, as the ``${\rm Pull_{SM}} $" reduces, these two leptoquarks are not effective in reducing the CC tension however they can enhance the $R_{K^{(*)}}^\nu$ contributions to the future scenario for $R_K^\nu$. Similarly for $V_2^\mu$, when comparing all three panels, we find it can produce the future scenario $R_K^\nu$ value with the couplings contributing insignificantly to NC anomalies.   In most cases the predictions for $R_K^\nu$ and $R_{K^*}^\nu$ overlap (red and blue points, respectively) as the contributing NP operator has SM-like (left-handed for both quark and neutrino currents) structure. The differences between the $K$ and $K^*$ modes are seen when we have a scalar, tensor or a right-handed quark current, which can easily be inferred from the expressions given in Eqs.~\eqref{eq:Kdist} and \eqref{eq:Kstdist}. We have also computed the prediction for $F_L^{K^*\nu\bar{\nu}}$ over the entire $q^2$ range and find that in most of the above mentioned cases the predictions coincide with the SM estimates except for $\tilde{R}_2$ and scenario $S_1$-III. In these two cases tensor operators are involved, which then reduce the $F_L^{K^*\nu\bar{\nu}}$ value by a factor of 2 compared to the SM estimate.

\section{Summary}
\label{sec:summary}

In this article we explore the potential impact of a very recent Belle II measurement of $B^+\to K^+ \nu \bar{\nu}$ decay on several beyond the SM theories. The current world average of the branching fraction ($ (1.1\pm 0.4)\times 10^{-5}$) including the new Belle II result indicates a signal strength two times larger than the SM prediction although with large uncertainty. However, a reanalysis of the full Belle dataset with the new tagging method and future Belle II results can significantly reduce the uncertainty. The $B\to K \nu \bar{\nu}$ mode is free from nonfactorizable contributions compared to the charged lepton channel $B\to K^{(*)}\ell\ell$ and thus any deviation from the SM prediction of the branching ratio constitutes a clear signal of NP.
In view of the expected future improvement in data, we present two very popular NP scenarios namely leptoquarks and generic $Z^\prime$ models that contribute to the $b\to s \nu \bar{\nu} $ process via tree level interactions. The choice of NP scenarios is motivated by the possibility of addressing either or both of the NC and CC discrepancies observed in $B$-decays by several experimental collaborations. 

In order to find a correlation between the $b\to s \mu \mu$, $b\to c \tau \bar{\nu}$ and $b\to s \nu \bar{\nu}$ modes, we include light RHNs in our analysis. Starting with the most general dimension-6 beyond the SM Hamiltonian, we derive the differential branching fractions with respect to the dineutrino invariant mass squared for $B\to K^{(*)} \nu \bar{\nu}$ modes and the longitudinal helicity fraction of the $K^*$ for the vector boson mode. The effect of NP mediators is captured by four fermion operators that are obtained by matching with the SMEFT as well as the $\nu$-WET basis. Apart from the (axial)vector operators, the RHNs induce scalar and tensor structures and have interesting consequences. Focusing on a minimal set of NP couplings we find that the scalar leptoquark $S_1$ can produce the $B^+\to K^+ \nu \bar{\nu}$ signal strength and also explain the CC anomalies within their $\pm 1 \sigma$ uncertainties. In terms of its combined contribution to CC and NC sectors, the scalar leptoquark $S_3$ is a good candidate although it barely reduces the tensions in $R(D^{(*)})$, its effects can increase the $B^+\to K^+ \nu \bar{\nu}$ contribution to the desired range. We find that the vector leptoquark $U_3$ can significantly enhance  the $B^+\to K^+ \nu \bar{\nu}$ rate and also explain the NC anomalies to a great extent. Interestingly, addressing $b\to c \tau \bar{\nu}$ tensions in scenarios with RHNs, due to the absence of interference with the SM contributions, the required large couplings enhance $b\to s \nu \bar{\nu} $ by 2 orders of magnitude, and hence are completely excluded.  For the vector boson $Z^\prime$, we infer that with the minimal set of parameters that can explain the NC anomalies, $B^+\to K^+ \nu \bar{\nu}$ remains very close to the SM expectations.

While this article focuses on exploring the connection of $B^+\to K^+ \nu \bar{\nu}$ with $B$-anomalies in specific NP models, a discussion of other NP theories e.g., models with a dark matter candidate is left for our future work. In such cases, apart from the branching ratios, the $q^2$ variation of angular distributions and observables such as the helicity fraction of the $K^*$ in the $B\to K^* \nu \bar{\nu}$ mode would also be interesting to study for the discrimination of different NP models.

\subsubsection*{Acknowledgment}
We thank Alexander Glazov for useful comments. R.M. thanks Ayan Paul, Marco Fedele and Mauro Valli for useful discussions. The work of R.M. is partly supported by the Alexander von Humboldt Foundation and the University of Siegen under the Young Investigator Research Group (Nachwuchsforscherinnengruppe) grant. T.E.B. acknowledges support from the U.S. Department of Energy (DOE) Office of High Energy Physics (OHEP) Award No. DE-SC0010504.

\appendix

\section{Helicity amplitudes}
\label{sec:FF}

Here, we give the expressions for the helicity amplitudes used in Eqs.~\eqref{eq:Kdist} and \eqref{eq:Kstdist} in terms of the form factors. We start with the $B \to K$ transition:
\bea
\nn
H_{V}^s (q^2) = \,  \sqrt{\frac{\lambda_K(q^2)}{q^2}}\, f_+(q^2) \, , ~~
H_S^s(q^2) =  \frac{m_B^2 - m_K^2}{m_b - m_s}\, f_0(q^2) \, , ~~
H_T^s(q^2) = \,  - \frac{\sqrt{\lambda_K(q^2)}}{m_B+m_K}\, f_T(q^2) \, ,
\label{eq:hadampD}
\eea
and for $B \to K^*$:
\bea
\nn
H_{V, \pm} (q^2) &=& (m_B+m_{K^*})\, A_1(q^2) \mp \frac{\sqrt{\lambda_{K^*} (q^2)}}{m_B + m_{K^*}}\, V(q^2) \, , 
\\ \nn
H_{V,0} (q^2) 
&=& \frac{8}{ \sqrt{q^2}}\, m_B m_{K^*}  \, A_{12}(q^2) , 
\\ 
H_S(q^2) &=& - \frac{\sqrt{\lambda_{K^*}(q^2)}}{m_b+m_c}\, A_0(q^2)\, , 
\\ \nn
H_{T,0}(q^2) &=&- \frac{4m_B m_{K^*}}{m_B+ m_{K^*}} T_{23}(q^2) , \\ \nn 
H_{T\pm}(q^2) &=&  \frac{1}{\sqrt{q^2}}\left[\pm (m_B^2-m_{K^*}^2)\, T_2(q^2) + \sqrt{\lambda_{K^*}}\, T_1(q^2)\right] .
\label{eq:hadampDstar}                          
\eea
The inputs for the form factors $f_i(q^2)$ are taken from lattice QCD computations~\cite{Bailey:2015dka} and $V,A_{1,12},T_{1,2,23}$ are from a combined LCSR and lattice QCD analysis~\cite{Straub:2015ica}.

\section{Operator Basis and QCD running}
\label{sec:SMEFT}

We list the four fermion semileptonic dimension-6 operators built from the SM fields written in the Warsaw basis~\cite{Grzadkowski:2010es}.
\begin{align}
\label{eq:SMEFTop}
[\mathcal{O}_{lq}^{(3)}]^{ij\alpha\beta}=&(\bar{Q}^i \gamma^\mu \sigma^a Q^j)(\bar{L}^\alpha \gamma_\mu \sigma^a L^\beta)\,, &
[\mathcal{O}_{lq}^{(1)}]^{ij\alpha\beta}=&(\bar{Q}^i \gamma^\mu Q^j)(\bar{L}^\alpha \gamma_\mu  L^\beta) \,,&\\
[\mathcal{O}_{ld}]^{ij\alpha\beta}=&(\bar{d}^i_R\gamma^\mu d_R^j)(\bar{L}^\alpha\gamma_\mu L^\beta) \,,& 
[\mathcal{O}_{lu}]^{ij\alpha\beta}=&(\bar{u}^i_R\gamma^\mu u_R^j)(\bar{L}^\alpha\gamma_\mu L^\beta) \,,& \\
[\mathcal{O}_{eq}]^{ij\alpha\beta}=&(\bar{Q}^i\gamma^\mu Q^j)(\bar{e}^\alpha_R\gamma_\mu e^\beta_R) \,,&
[\mathcal{O}_{eu}]^{ij\alpha\beta}=&(\bar{u}_R^i\gamma^\mu u_R^j)(\bar{e}^\alpha_R\gamma_\mu e^\beta_R) \,,& \\
[\mathcal{O}_{ed}]^{ij\alpha\beta}=&(\bar{d}_R^i\gamma^\mu d_R^j)(\bar{e}^\alpha_R\gamma_\mu e^\beta_R) \,,& 
[\mathcal{O}_{ledq}]^{ij\alpha\beta}=&(\bar{d}_R^i Q^j)(\bar{L}^\alpha  e_R^\beta) \,,& \\
[\mathcal{O}_{lequ}^{(1)}]^{ij\alpha\beta}=&(\bar{Q}^i u_R^j)\epsilon(\bar{L}^\alpha e^\beta_R) \,,&\
[\mathcal{O}_{lequ}^{(3)}]^{ij\alpha\beta}=&(\bar{Q}^i \sigma^{\mu\nu}u_R^j)\epsilon(\bar{L}^\alpha\sigma_{\mu\nu} e^\beta_R) \,.& 
\end{align}
We denote the left-handed SM quark (lepton) doublets as $Q$ ($L$), while $u_R$ ($d_R$) and $\ell_R$ are the right-handed up-type (down-type) quark and lepton singlets, respectively. Here $\epsilon = i\sigma_2$ is the antisymmetric isospin tensor. We adopt a `down' basis in which the down-type-quark and charged-lepton Yukawa matrices are diagonal. In this basis, the transformation from the fermion interaction eigenstates to mass eigenstates is simply given by $u_L \to V^\dagger u_L$ where $V$ is the quark CKM matrix.

While inclusion of the light SM gauge singlet RHN $\nu_R(1,1,0)$ induces a few more operators, these of course cannot be cast in the SMEFT basis, and at energies below the electroweak scale it is convenient to use the $\nu$-WET, yielding for neutral current transitions
\begin{align}
\label{eq:WET}
\mathcal{H}_{\mathrm{eff}}^{\mathrm{nc},\nu}\; = \;- \sum_{q=u,d} &  \bigg\{
[N_{LL}^{V,q}]^{ij\alpha\beta}\, (\bar q_L^i\gamma^\mu q_L^j) (\bar \nu_L^\alpha\gamma_\mu \nu_L^\beta) 
\, +\, 
[N_{RL}^{V,q}]^{ij\alpha\beta}\, (\bar q_R^i\gamma^\mu q_R^j) (\bar \nu_L^\alpha\gamma_\mu \nu_L^\beta) \nn \\
& + [N_{LR}^{V,q}]^{ij\alpha\beta}\, (\bar q_L^i\gamma^\mu q_L^j) (\bar \nu_R^\alpha\gamma_\mu \nu_R^\beta) 
\, +\, 
[N_{RR}^{V,q}]^{ij\alpha\beta}\, (\bar q_R^i\gamma^\mu q_R^j) (\bar \nu_R^\alpha\gamma_\mu \nu_R^\beta) \nn \\
&+ [N_{LL}^{S,q}]^{ij\alpha\beta}\, (\bar q_R^i q_L^j) (\bar \nu_R^\alpha \nu_L^\beta) 
\, +\, 
[N_{RL}^{S,q}]^{ij\alpha\beta}\, (\bar q_L^i q_R^j) (\bar \nu_R^\alpha \nu_L^\beta) \nn \\
& + [N_{LR}^{S,q}]^{ij\alpha\beta}\, (\bar q_R^i q_L^j) (\bar \nu_L^\alpha \nu_R^\beta) 
\, +\, 
[N_{RR}^{S,q}]^{ij\alpha\beta}\, (\bar q_L^i q_R^j) (\bar \nu_L^\alpha \nu_R^\beta) \nn \\
& +[N_{LL}^{T,q}]^{ij\alpha\beta}\, (\bar q_L^i\sigma^{\mu\nu} q_L^j) (\bar \nu_L^\alpha\sigma_{\mu\nu} \nu_L^\beta) 
\, +\, 
[N_{RR}^{T,q}]^{ij\alpha\beta}\, (\bar q_R^i\sigma^{\mu\nu} q_R^j) (\bar \nu_R^\alpha\sigma_{\mu\nu} \nu_R^\beta) 
\bigg\}\, .
\end{align}

We recall that while considering heavy NP mediators such as leptoquarks, after performing the matching of operators to the above mentioned basis at the NP mass scale $\mu = M_{\mathrm{\tiny LQ}}$, we need to evolve them to the hadronic decay scale using the renormalization group equations. Neglecting electroweak corrections, we obtain
\be 
[N_{MN}^{X,q}]^{ij\alpha\beta}(\mu) \, =\, \Omega_N(\mu,M_{\mathrm{LQ}})\; [N_{MN}^{X,q}]^{ij\alpha\beta}(M_{\mathrm{LQ}})\, ,
\ee
where at the lowest order (leading logarithm), the evolution operator is given by
\be 
\Omega_N(\mu,M_{\mathrm{LQ}})\, =\, \left(\frac{\alpha_s^{(n_f)}(\mu)}{\alpha_s^{(n_f)}(m_q^{f+1})}\right)^{-\gamma^J_1/\beta_1^{(n_f)}}
\cdots\quad
\left(\frac{\alpha_s^{(5)}(m_b)}{\alpha^{(5)}_s(m_t)}\right)^{-\gamma^J_1/\beta_1^{(5)}}\;
\left(\frac{\alpha_s^{(6)}(m_t)}{\alpha^{(6)}_s(M_{\mathrm{LQ}})}\right)^{-\gamma^J_1/\beta_1^{(6)}},
\ee
with $n_f$ the relevant number of quark flavors at the hadronic scale considered and $m_q^{f+1}$ being the lightest (integrated-out) quark. The powers are governed by the first coefficients of the QCD $\beta$-function, $\beta_1^{(n_f)}= (2 n_f-33)/6$, and the anomalous dimensions for the currents are
\be 
\gamma_1^V = 0\, ,\qquad\qquad
\gamma_1^S = 2\, ,\qquad\qquad
\gamma_1^T = -2/3\, .
\ee
Note that the vector currents are not affected, while the scalar and tensor currents renormalize multiplicatively.

\section{Charged current observables}
\label{sec:CCexps}

The numerical expressions for  the observables $R(D^{(*)})$ and $F_L^{D^*}$ can be written as~\cite{Mandal:2020htr}
\begin{eqnarray}
\label{eq:RDRDSM}
{\cal R}({D}) / {\cal R}_({D})_\text{SM} &\!\! \!\!\approx &\!\!\!\! \left(|1+\mathcal{P}^V_{LL} + \mathcal{P}^V_{RL}|^2+|\mathcal{P}^V_{LR}+ \mathcal{P}^V_{RR}|^2\right) \nn
\\[3pt] \nn  &\!\!\! \!+ &\!\!\!\! \! 1.037 \left(|\mathcal{P}^S_{LL}+\mathcal{P}^S_{RL}|^2+|\mathcal{P}_{LR}^S+\mathcal{P}^S_{RR}|^2\right) +
0.939 \left(|\mathcal{P}^{T}_{LL}|^2+|\mathcal{P}^T_{RR}|^2\right) \nn
\\[3pt] \nn  &\!\! \!\!+ &\!\!\!\!\!  1.171\, \Re \left[ (1+\mathcal{P}^V_{LL} + \mathcal{P}_{RL}^{V})\, \mathcal{P}^{T*}_{LL} + (\mathcal{P}^V_{LR} + \mathcal{P}_{RR}^{V})\, \mathcal{P}^{T*}_{RR} \right] 
\\[3pt]  &\!\!\! \!+ &\!\!\!\!\! 
1.504 \,  \Re \left[(1+\mathcal{P}^V_{LL} + \mathcal{P}_{RL}^{V}) (\mathcal{P}^{S*}_{LL} +  \mathcal{P}^{S*}_{RL}) + (\mathcal{P}^V_{LR} + \mathcal{P}_{RR}^{V}) (\mathcal{P}^{S*}_{LR}\! +\!  \mathcal{P}^{S*}_{RR} ) \right], 
 \end{eqnarray}
 \begin{eqnarray}
{\cal R}({D^*}) / {\cal R}({D^*})_\text{SM} &\!\! \!\!\approx &\!\!\!\! \left(|1+\mathcal{P}^V_{LL}|^2+|\mathcal{P}^V_{RL}|^2+|\mathcal{P}^V_{LR}|^2+|\mathcal{P}^V_{RR}|^2\right) \nn
\\[3pt] \nn  &\!\!\! \!+ &\!\!\!\! \! 0.037 \left(|\mathcal{P}^S_{RL}-\mathcal{P}^S_{LL}|^2+|\mathcal{P}_{RR}^S-\mathcal{P}^S_{LR}|^2\right) \nn
\\[3pt] \nn  &\!\!\! \!+ &\!\!\!\! \!
17.378 \left(|\mathcal{P}^{T}_{LL}|^2+|\mathcal{P}^T_{RR}|^2\right) -1.781 \,  \Re \left[(1+\mathcal{P}^V_{LL})\, \mathcal{P}_{RL}^{V*}+\mathcal{P}^V_{LR}\, \mathcal{P}^{V*}_{RR}\right]
\\[3pt] \nn &\!\!\! \!+ &\!\!\!\! \!
5.748 \, \Re \left[ \mathcal{P}_{RL}^V \mathcal{P}^{T*}_{LL} +\mathcal{P}^V_{LR} \mathcal{P}^{T*}_{RR} \right] - 5.130 \, \Re \left[(1+\mathcal{P}^V_{LL})\, \mathcal{P}^{T*}_{LL} +\mathcal{P}^V_{RR}\, \mathcal{P}^{T*}_{RR}\right]
\\[3pt]  &\!\!\! \!+ &\!\!\!\! \!
0.114 \, \Re \left[ (1+\mathcal{P}^V_{LL}\!-\!\mathcal{P}_{RL}^V)\, (\mathcal{P}^{S*}_{RL}-\mathcal{P}^{S*}_{LL})+(\mathcal{P}^V_{RR}\!-\!\mathcal{P}^V_{LR})\, (\mathcal{P}_{LR}^{S*}\!-\!\mathcal{P}^{S*}_{RR})\right],
\label{eq:RDstRDSM} \\[3ex]
F_L^{D^*}\! \times {\cal R}({D^*}) &\!\!\!\!\approx&\!\!\!\! 0.120 \left( 
|1+\mathcal{P}_{LL}^V-\mathcal{P}_{RL}^V|^2     
+|\mathcal{P}_{RR}^V-\mathcal{P}_{LR}^V|^2\right) \nn
\\[3pt] \nn &\!\!\! \!+ &\!\!\!\! \!
0.010 \left( |\mathcal{P}^S_{RL} - \mathcal{P}^S_{LL}|^2 + |\mathcal{P}^S_{RR} - \mathcal{P}_{LR}^S|^2\right)
+ 0.869 \left( |\mathcal{P}^{T}_{LL}|^2 + |\mathcal{P}^T_{RR} |^2 \right)
\\[3pt] \nn &\!\!\! \!+ &\!\!\!\! \!
0.030\, \Re \left[ (1+\mathcal{P}_{LL}^V-\mathcal{P}_{RL}^V)     
\, (\mathcal{P}^{S*}_{RL} - \mathcal{P}^{S*}_{LL}) - (\mathcal{P}^V_{RR} - \mathcal{P}^V_{LR})\, (\mathcal{P}_{RR}^{S*} - \mathcal{P}^{S*}_{LR})\right]  
\\[3pt] &\!\!\! \!+ &\!\!\!\! \!
0.525 \, \Re \left[ (1+\mathcal{P}_{LL}^V-\mathcal{P}_{RL}^V)     
\, \mathcal{P}^{T*}_{LL}+ (\mathcal{P}^V_{RR} -\mathcal{P}^V_{LR} )\, \mathcal{P}^{T*}_{RR} \right] .
\label{eq:FLDRDst}
\end{eqnarray}
Note that the Wilson coefficients $\mathcal{P}^X_{AB}$ are evaluated at the $m_b$-scale where we have dropped `33' superscripts from all of them for simplicity. The running from a NP scale $\Lambda$ to $m_b$ can be incorporated using renormalization group equations, and neglecting electroweak contributions we obtain for the scalar and tensor operators 
$$\mathcal{P}^{S(T)}_{AB}(m_b) = 1.67(0.84)\times \mathcal{P}^{S(T)}_{AB} (\Lambda=\mathcal{O}({\rm TeV}))\,.$$

 In Eqs.~\eqref{eq:RDRDSM}-\eqref{eq:FLDRDst}, NP is assumed only in the third generation of leptons and the form factors follow the Boyd, Grinstein and Lebed parametrization~\cite{Boyd:1994tt} in the heavy quark effective theory including corrections of order $\alpha_s$, $\Lambda_{\rm QCD}/m_{b,c}$~\cite{Bernlochner:2017jka} and partly $\Lambda_{\rm QCD}^2/m_{c}^2$~\cite{Jung:2018lfu}.

\end{document}